\documentclass[onecolumn]{elsart3p}
\usepackage{amssymb}
\usepackage{psfrag}
\usepackage{graphicx}
\usepackage{alltt}
\usepackage{natbib}
\usepackage[fleqn]{amsmath}
\usepackage{color}
\usepackage{colortbl}
\journal{Boundary-Layer Meteorology}
\newcommand{\bv}[1]{{\mbox{\boldmath$#1$}}}
\newcommand{\mean}[1]{{\left<#1\right>}}

\newcommand*{\Eqr}[1]{(\ref{#1})}
\newcommand*{\Eqre}[1]{Equation~(\ref{#1})}
\newcommand*{\Eqrs}[1]{(\ref{#1})}
\newcommand*{\Eqres}[1]{Equations~(\ref{#1})}

\newcommand*{\Fige}[1]{Figure~\ref{#1}}

\begin{document}

\begin{frontmatter}
\title{\textbf{Joint PDF modelling of turbulent flow and dispersion in an urban street canyon}}
\author{J.\ Bakosi\corauthref{cor}}
\corauth[cor]{Corresponding author.}
\address{Los Alamos National Laboratory, Los Alamos, NM 87545}
\ead{jbakosi@lanl.gov}
\author{P.\ Franzese}
\author{and Z.\ Boybeyi}
\address{College of Science, George Mason University, Fairfax, VA 22030}
\begin{abstract}
The joint probability density function (PDF) of turbulent velocity and concentration of a passive scalar in an urban street canyon is computed using a newly developed particle-in-cell Monte Carlo method. Compared to moment closures, the PDF methodology provides the full one-point one-time PDF of the underlying fields containing all higher moments and correlations. The small-scale mixing of the scalar released from a concentrated source at the street level is modelled by the interaction by exchange with the conditional mean (IECM) model, with a micro-mixing time scale designed for geometrically complex settings. The boundary layer along no-slip walls (building sides and tops) is fully resolved using an elliptic relaxation technique, which captures the high anisotropy and inhomogeneity of the Reynolds stress tensor in these regions. A less computationally intensive technique based on wall functions to represent boundary layers and its effect on the solution are also explored. The calculated statistics are compared to experimental data and large-eddy simulation. The present work can be considered as the first example of computation of the full joint PDF of velocity and a transported passive scalar in an urban setting. The methodology proves successful in providing high level statistical information on the turbulence and pollutant concentration fields in complex urban scenarios.
\end{abstract}
\end{frontmatter}
\section{Introduction}
Regulatory bodies, architects and town planners increasingly use computer models to assess ventilation and occurrences of hazardous pollutant concentrations in cities. These models are mostly based on the Reynolds-averaged Navier-Stokes (RANS) equations or, more recently, large-eddy simulation (LES) techniques. Both of these approaches require a series of modelling assumptions, including most commonly the eddy-viscosity and gradient-diffusion hypotheses. The inherent limitations of these approximations, even in the simplest engineering flows, are well known, and detailed for example by \cite{Pope_00}. Although the effects of the assumptions are more pronounced in RANS than in LES models, where they are confined to the smaller (modelled) scales, there is clearly a need to develop higher-order models. In pollutant dispersion modelling it is also desirable to predict extreme events such as peak values or probabilities that concentrations will exceed a certain threshold. In other words, a fuller statistical description of the concentration field is required \citep{Chatwin_93,Kristensen_94,Wilson_95,Pavageau_99}. These issues have been explored in grid turbulence and in the unobstructed atmosphere, and models capable of predicting higher-order statistics have also appeared \citep{Yee_et_al_94,Yee_Wilson_00,Luhar_et_al_00,Cassiani_Giostra_02,Franzese_03,Cassiani_et_al_05,Cassiani_et_al_05b}, but more research is necessary to extend these capabilities to cases of built-up areas.

Probability density function (PDF) methods have been developed mainly within the combustion engineering community as an alternative to moment closure techniques to simulate chemically reactive turbulent flows \citep{Lundgren_69,Pope_85,Dopazo_94}. Because many-species chemistry is high-dimensional and highly non-linear, the biggest challenge in reactive flows is to adequately model the chemical source term. In PDF methods, the closure problem is raised to a statistically higher level by solving for the full PDF of the turbulent flow variables instead of its moments. This has several benefits: convection, the effect of mean pressure, viscous diffusion and chemical reactions appear in closed form in the PDF transport equation. Therefore these processes are treated mathematically exactly without closure assumptions, eliminating the need for gradient-transfer approximations. The effects of fluctuating pressure, dissipation of turbulent kinetic energy and small-scale mixing of scalars still have to be modelled. The rationale is that, since the most important physical processes are treated exactly, the errors introduced by modelling assumptions for less important processes amount to a smaller departure from reality. Moreover, the higher level description provides more information that can be used in the construction of closure models.

The PDF transport equation is a high-dimensional scalar equation. Although techniques of solution based on stochastic Eulerian methods have been developed \citep{Valino_98,Mobus_01,Soulard_06}, Lagrangian Monte Carlo methods are a more natural choice because their computational cost increases only linearly with the dimensionality of the problem, favourably comparing to the more traditional finite difference, finite volume or finite element methods.

The numerical development of Lagrangian PDF methods has mainly centred around three distinctive approaches, all of them representing a finite ensemble of fluid particles with Lagrangian particles. A common approach is the \emph{stand-alone Lagrangian} method, where the flow is represented by particles whereas the Eulerian statistics are obtained using kernel estimation \citep{Pope_00,Fox_03}. Another technique is the \emph{hybrid} methodology, which builds on existing Eulerian computational fluid dynamics (CFD) codes based on moment closures \citep{Muradoglu_99,Muradoglu_01,Jenny_01,Rembold_06}. Hybrid methods use particles to solve for certain quantities and provide closures for the Eulerian moment equations using the particle/PDF methodology. A more recent approach is the self-consistent \emph{non-hybrid} method \citep{Bakosi_07,Bakosi_08}, which also employs particles to represent the flow, and uses the Eulerian grid only to solve for inherently Eulerian quantities (such as the mean pressure) and for efficient particle tracking. Since the latter two approaches extensively employ Eulerian grids, they are particle-in-cell methods \citep{Grigoryev_02}.

The current study presents an application of the non-hybrid method to a simplified urban-scale case where pollution released from a concentrated line source between idealized buildings is simulated and results are compared to data from wind-tunnel experiments and LES.

PDF methods in atmospheric modelling mostly focus on the simulation of passive pollutants, wherein the velocity field (mean and turbulence) is either assumed or obtained from experiments \citep{Sawford_04,Sawford_06,Cassiani_et_al_05,Cassiani_et_al_05b,Cassiani_07}. In contrast, the current model directly computes the joint PDF of the turbulent velocity, characteristic turbulent frequency and scalar concentration, extending the use of PDF methods in atmospheric modelling to represent more physics at a higher statistical level. A computed full joint PDF also has the advantage of providing information on the uncertainty originating from turbulence on a physically sound basis.

In this study the turbulent boundary layers developing along solid walls are treated in two different ways: either fully resolved or via the application of wall functions (i.e.\ the logarithmic ``law of the wall''). The full resolution is obtained using Durbin's elliptic relaxation technique \citep{Durbin_93}, which was incorporated into the PDF methodology by \citet{Dreeben_97,Dreeben_98}. This technique allows for an adequate representation of the near-wall low-Reynolds-number effects, such as the high inhomogeneity and anisotropy of the Reynolds stress tensor and wall-blocking. Wall conditions for particles based on the logarithmic ``law of the wall'' in the PDF framework have also been developed by \citet{Dreeben_97b}. These two types of wall treatments are examined in terms of trade-off between computational cost and performance, addressing the question of how important it is to adequately resolve the boundary layers along solid walls in order to obtain reasonable scalar statistics.

At the urban scale the simplest settings to study turbulent flow and dispersion patterns are street canyons. Due to increasing concerns for environmental issues and air quality standards in cities, a wide variety of canyon configurations and release scenarios have been studied both experimentally \citep{Hoydysh_74,Wedding_77,Rafailids_95,Meroney_96,Pavageau_99} and numerically \citep{Lee_94,Johnson_95,Baik_99,Huang_00,Liu_02}. Street canyons have a simple flow geometry, they can be studied in two dimensions, and a wealth of experimental and modelling data are available for different street-width to building-height ratios. This makes them ideal candidates for testing a new urban pollution dispersion model. We validate the computed velocity and scalar statistics with the LES results of \citet{Liu_02} and the wind-tunnel measurements of \citet{Meroney_96,Pavageau_96,Pavageau_99}. The experiments have been performed in the atmospheric wind tunnel of the University of Hamburg, where the statistics of the pollutant concentration field have been measured in an unusually high number of locations in order to provide fine details inside the street canyon.

The rest of the paper is organized as follows. In Section \ref{sec:governing_equations}, the exact and modelled governing equations are presented. Several statistics are compared to experimental data and large-eddy simulations in Section \ref{sec:street_canyon}. Finally, Section \ref{sec:discussion} draws some conclusions and elaborates on possible future directions.
\section{Governing equations}
\label{sec:governing_equations}
We write the Eulerian governing equations for a passive scalar released in a viscous, Newtonian, incompressible flow as
\begin{align}
\frac{\partial U_i}{\partial x_i} &= 0,\\
\frac{\partial U_i}{\partial t} + U_j\frac{\partial U_i}{\partial x_j} + \frac{1}{\rho}\frac{\partial P}{\partial x_i} &=  \nu\nabla^2U_i \label{eq:NavierStokes},\\
\frac{\partial\phi}{\partial t} + U_i\frac{\partial\phi}{\partial x_i} &= \Gamma\nabla^2\phi,\label{eq:scalar}
\end{align}
where $U_i$, $P$, $\rho$, $\nu$, $\phi$ and $\Gamma$ are the Eulerian velocity, pressure, constant density, kinematic viscosity, scalar concentration and scalar diffusivity, respectively. Based on this system of equations the exact transport equation that governs the one-point, one-time Eulerian joint PDF of velocity and concentration $f(\bv{V},\psi;\bv{x},t)$ can be written as \citep{Pope_85,Pope_00},
\begin{equation}
\begin{split}
\frac{\partial f}{\partial t} + V_i\frac{\partial f}{\partial x_i} &= \nu\frac{\partial^2f}{\partial x_i\partial x_i} + \frac{1}{\rho}\frac{\partial\mean{P}}{\partial x_i}\frac{\partial f}{\partial V_i} - \frac{\partial^2}{\partial V_i\partial V_j}\left(f\mean{\nu\frac{\partial U_i}{\partial x_k}\frac{\partial U_j}{\partial x_k}\Bigg|\bv{U}=\bv{V},\phi=\psi}\right)\\
&\quad+ \frac{\partial}{\partial V_i}\left(f\mean{\frac{1}{\rho}\frac{\partial p}{\partial x_i}\Bigg|\bv{U}=\bv{V},\phi=\psi}\right) - \frac{\partial}{\partial\psi}\Big(f\big<\Gamma\nabla^2\phi\big|\bv{U}=\bv{V},\phi=\psi\big>\Big),
\end{split}
\label{eq:EulerianPDFforviscous}
\end{equation}
where $\bv{V}$ and $\psi$ denote the sample space variables of the stochastic velocity $\bv{U}(\bv{x},t)$ and concentration $\phi(\bv{x},t)$ fields, respectively, and the pressure $P$ is decomposed into its mean $\mean{P}$ and fluctuation part $p$.  In \Eqre{eq:EulerianPDFforviscous} the physical processes of advection (second term on the left), viscous diffusion (first term on the right) and transport of $f$ in velocity space by the mean pressure gradient (second term on the right) are represented mathematically exactly. The last three terms in the form of conditional expectations have to be modelled: these are respectively the effect of dissipation of turbulent kinetic energy, the effect of fluctuating pressure and the small-scale diffusion of the scalar. After appropriate modelling of the unclosed terms, \Eqre{eq:EulerianPDFforviscous} can, in principle, be solved with a traditional numerical method. However, the high-dimensionality makes Monte Carlo methods more appealing. In particular, because \Eqre{eq:EulerianPDFforviscous} is a Fokker-Planck equation, it can be written as an equivalent system of stochastic differential equations (SDEs) \citep{vanKampen_04}. We use the generalized Langevin model (GLM) of \citet{Haworth_86} for the velocity increment, and the interaction by exchange with the conditional mean (IECM) model for the scalar. The physics and characteristics of the IECM model are discussed in detail by \citet{Fox_96,Pope_98} and \citet{Sawford_04}. Thus our system of SDEs that solve \Eqre{eq:EulerianPDFforviscous} is written as
\begin{eqnarray}
\mathrm{d}\mathcal{X}_i&=&\mathcal{U}_i\mathrm{d}t + \left(2\nu\right)^{1/2}\mathrm{d}W_i,\label{eq:Lagrangian-position}\\
\mathrm{d}\mathcal{U}_i&=&-\frac{1}{\rho}\frac{\partial\mean{P}}{\partial x_i}\mathrm{d}t + 2\nu\frac{\partial^2\mean{U_i}}{\partial x_j\partial x_j}\mathrm{d}t + \left(2\nu\right)^{1/2}\frac{\partial\mean{U_i}}{\partial x_j}\mathrm{d}W_j+G_{ij}\left(\mathcal{U}_j-\mean{U_j}\right)\mathrm{d}t + \left(C_0\varepsilon\right)^{1/2}\mathrm{d}W'_i,\label{eq:Lagrangian-velocity}\\
\mathrm{d}\psi&=&-\frac{1}{t_\mathrm{m}}\left(\psi-\mean{\phi|\bv{U}=\bv{V}}\right)\mathrm{d}t.\label{eq:IECM}
\end{eqnarray}
\Eqre{eq:Lagrangian-position} governs the Lagrangian particle position $\mathcal{X}_i$, and it consists of advection by the instantaneous particle velocity $\mathcal{U}_i$ and molecular diffusion represented by the isotropic Wiener process $\mathrm{d}W_i$, which is a given Gaussian process with zero mean and variance $\mathrm{d}t$. The particle velocity $\mathcal{U}_i$ is governed by \Eqre{eq:Lagrangian-velocity}. The second and third terms on the right-hand side are a direct consequence of the particular Lagrangian representation of the viscous diffusion in \Eqre{eq:Lagrangian-position} \citep[for a derivation, see][]{Dreeben_97}, thus the first three terms on the right-hand side govern the particle velocity in a laminar flow mathematically exactly. The last two terms involving the tensor $G_{ij}$ and $C_0$ jointly model the effect of pressure redistribution and anisotropic dissipation of turbulent kinetic energy. $G_{ij}$ is a second-order tensor function of the mean velocity gradients $\partial\mean{U_i}/\partial x_j$, the Reynolds stresses $\mean{u_iu_j}$ and the rate of dissipation of turbulent kinetic energy $\varepsilon$, while $C_0$ is a positive constant. Note that the Wiener process $\mathrm{d}W_i$ appearing in both \Eqres{eq:Lagrangian-position} and \Eqr{eq:Lagrangian-velocity} is the same process, i.e.\ the same exact series of random numbers, and is independent of the other process $\mathrm{d}W_i'$. The concentration of the passive scalar is governed by \Eqre{eq:IECM}, which represents the physical process of diffusion by relaxation of the particle scalar $\psi$ towards the velocity-conditioned scalar mean $\mean{\phi|\bv{U}=\bv{V}}\equiv\mean{\phi|\bv{V}}$ with a timescale $t_\mathrm{m}$.  Note that the Eulerian statistics denoted by angled brackets $\mean{\cdot}$ are to be evaluated at the fixed particle locations $\mathcal{X}_i$, and in particle-in-cell methods, this is usually achieved using an Eulerian grid and computing ensemble averages in grid elements and/or around vertices. 
The energy dissipation rate is defined as
\begin{equation}
\varepsilon = \mean{\omega}\left(k + \nu C_T^2\mean{\omega}\right),\label{eq:dissipation-from-frequency_wall}
\end{equation}
where $C_T$ is a model constant.  We adopt the model of \citet{VanSlooten_98} for the stochastic characteristic turbulent frequency $\omega$
\begin{equation}
\mathrm{d}\omega = -C_3\mean{\omega}\big(\omega-\mean{\omega}\big)\mathrm{d}t - S_\omega\mean{\omega}\omega\mathrm{d}t+\big(2C_3C_4\mean{\omega}^2\omega\big)^{1/2}\mathrm{d}W,\label{eq:frequency-model}
\end{equation}
where $S_\omega$ is a source/sink term for the mean turbulent frequency
\begin{equation}
S_\omega=C_{\omega2}-C_{\omega1}\frac{\mathcal{P}}{\varepsilon},\label{eq:frequency_source}
\end{equation}
where $\mathcal{P}=-\mean{u_iu_j}\partial\mean{U_i}/\partial x_j$ is the production of turbulent kinetic energy, $\mathrm{d}W$ is a scalar valued Wiener process, while $C_3,C_4,C_{\omega1}$ and $C_{\omega2}$ are model constants.  To define the micro-mixing time scale for a scalar released from a concentrated source in a geometrically complex flow domain we follow \citet{Bakosi_07,Bakosi_08} and specify the inhomogeneous $t_\mathrm{m}$ as
\begin{equation}
t_\mathrm{m}(\bv{x}) = \min\left[C_s\left(\frac{r_0^2}{\varepsilon}\right)^{1/3} + C_t\frac{|\bv{x}-\bv{x}_0|}{U_c(\bv{x})}; \enskip \max\left(\frac{k}{\varepsilon}; \enskip C_T\sqrt{\frac{\nu}{\varepsilon}}\right)\right],\label{eq:micromixing-timescale}
\end{equation}
where $r_0$ denotes the radius of the source, $U_c$ is a characteristic velocity at $\bv{x}$ that we take as the Euclidean norm of the mean velocity vector, $\bv{x}_0$ is the location of the source, and $C_s$ and $C_t$ are model constants.
\subsection{Elliptic relaxation modelling of near-wall turbulence}
The turbulence model GLM (represented by the last two terms of \Eqre{eq:Lagrangian-velocity} for the velocity increments) is in fact a family of models. A specific definition of $G_{ij}$ and $C_0$ corresponds to a particular closure from a wealth of models, many of which can be made equivalent (at the level of second moments) to popular Reynolds-stress closures \citep{Pope_94}. To be able to capture the near-wall low-Reynolds-number effects on the Reynolds stresses in fully resolved boundary layers, we follow \citet{Durbin_93} and \citet{Dreeben_98} and specify $G_{ij}$ and $C_0$ through the tensor $\wp_{ij}$
\begin{eqnarray}
G_{ij} &=& \frac{2\wp_{ij}-\varepsilon\delta_{ij}}{2k},\label{eq:Gij}\\
C_0 &=& -\frac{2\wp_{ij}\mean{u_iu_j}}{3k\varepsilon},\label{eq:C0}
\end{eqnarray}
where $k=\frac{1}{2}\mean{u_iu_i}$ denotes the turbulent kinetic energy and $\wp_{ij}$ is obtained by solving the elliptic equation
\begin{equation}
\wp_{ij} - L^2\nabla^2\wp_{ij} = \tfrac{1}{2}(1-C_1)k\mean{\omega}\delta_{ij} + kH_{ijkl}\frac{\partial\mean{U_k}}{\partial x_l},\label{eq:elliptic-relaxation-Lagrangian}
\end{equation}
where the fourth-order tensor $H_{ijkl}$ is given by
\begin{equation}
H_{ijkl} = (C_2A_v + \tfrac{1}{3}\gamma_5)\delta_{ik}\delta_{jl} - \tfrac{1}{3}\gamma_5\delta_{il}\delta_{jk}+\gamma_5b_{ik}\delta_{jl} - \gamma_5b_{il}\delta_{jk},\label{eq:H}
\end{equation}
with
\begin{equation}
A_v = \min\left[1; \enskip C_v\left(\tfrac{2}{3}k\right)^{-3}\det\mean{u_iu_j}\right],
\end{equation}
the Reynolds stress anisotropy 
\begin{equation}
b_{ij} = \frac{\mean{u_iu_j}}{\mean{u_ku_k}} - \tfrac{1}{3}\delta_{ij},
\end{equation}
and $C_1, C_2, \gamma_5, C_v$ are model constants. The characteristic length scale $L$ in \Eqre{eq:elliptic-relaxation-Lagrangian} is defined by the maximum of the turbulent and Kolmogorov length scales
\begin{equation} 
L=C_L \max\left[C_\xi \big(k^{3/2}/\varepsilon\big); \enskip C_\eta\left(\nu^3/\varepsilon\right)^{1/4}\right],\label{eq:L}
\end{equation}
with $C_\xi=1 + 1.3n_in_i$, where $C_L$ and $C_\eta$ are model constants, and $n_i$ is the unit wall-normal of the closest wall element pointing outward of the flow domain. For the applied wall-boundary conditions in this fully-resolved case the reader is referred to \citet{Dreeben_98} and \citet{Bakosi_08}. More details on the inflow and outflow conditions for the mean pressure and on the wall conditions for the tensor $\wp_{ij}$ are given in \citep{Bakosi_08}. \Eqre{eq:elliptic-relaxation-Lagrangian} was developed in conjunction with turbulent channel flow \citep{Durbin_93,Dreeben_98}. Modifications and different forms of this idea have also been proposed \citep{Whizman_96,Dreeben_97,Dreeben_98,Waclawczyk_04}, and an application in channel flow with the current non-hybrid model is presented by \citet{Bakosi_07}.  The model to compute the joint PDF of velocity, scalar, and characteristic turbulent frequency is now complete. 
\subsection{Wall functions modelling of near-wall turbulence}
\label{sec:wall-functions}
When the wall region has sufficient resolution, $G_{ij}$ and $C_0$ as defined by \Eqres{eq:Gij} and \Eqr{eq:C0} enable the model to adequately capture the near-wall effects on the higher-order turbulence statistics. Full resolution at the wall is strictly required in certain cases such as, for example, computations of heat transfer at walls embedded in a flow or detaching boundary layers with high adverse pressure gradients. In many other realistic simulations, however, full resolution of high-Reynolds-number boundary layers is not always possible and may not be necessary when the flow details close to walls are not important because the analysis focuses on the boundary-layer effects at farther distances. In these cases an alternative option is to model the near-wall turbulence using wall functions instead of the elliptic relaxation technique. Employing wall functions for no-slip walls provides a trade-off between the accuracy of fully resolved boundary layers and computational speed.  Wall functions are widely applied in atmospheric simulations, where full wall resolution is usually prohibitively expensive even at the microscale or urban scale \citep{Bacon_00,Lien_04}. It is worth emphasizing that one of the main assumptions used in the development of wall functions is that the boundary layer remains attached, which is not always the case in simulations of complex flows. However, since wall functions are the only choice for realistic atmospheric simulations, they are still routinely employed with reasonable success.

To investigate the gain in performance and the effect on the results, we implemented the wall treatment for complex flow geometries that has been developed for the PDF method by \citet{Dreeben_97b}. In this case, the tensor $G_{ij}$ is defined by the simplified Langevin model (SLM) \citep{Haworth_86} and $C_0$ is simply a constant:
\begin{equation}
G_{ij} = -\left(\tfrac{1}{2}+\tfrac{3}{4}C_0\right)\mean{\omega}\delta_{ij},\label{eq:wall-GandC}
\end{equation}
with $C_0=3.5$. In line with the purpose of wall functions, boundary conditions have to be imposed on particles that hit the wall so that their combined effect on the statistics at the first grid point from the wall will be consistent with the universal logarithmic wall function in equilibrium flows, i.e.\ in boundary layers with no significant adverse pressure gradients. The development of boundary conditions based on wall functions rely on the self-similarity of attached boundary layers close to walls. These conditions are applied usually at the first grid point from the wall based on the assumption of constant or linear stress distribution. This results in the well-known self-similar logarithmic profile for the mean velocity. For the sake of completeness the conditions on the particles developed by \citet{Dreeben_97b} are repeated here. The condition for the wall-normal component of the particle velocity reads
\begin{equation}
\mathcal{V}_{\scriptscriptstyle R} = -\mathcal{V}_{\scriptscriptstyle I},\label{eq:wall-normal-velocity}
\end{equation}
where the subscripts $R$ and $I$ denote reflected and incident particle properties, respectively. The reflected streamwise particle velocity is given by
\begin{equation}
\mathcal{U}_{\scriptscriptstyle R} = \mathcal{U}_{\scriptscriptstyle I} + \alpha\mathcal{V}_{\scriptscriptstyle I},
\end{equation}
where the coefficient $\alpha$ is determined by imposing consistency with the logarithmic law at the distance of the first grid point from the wall, $y_p$:
\begin{equation}
\alpha = \frac{2\Hat{u}_p^2\mean{U}_p|\mean{U}_p|}{\mean{v^2}_pU_e^2},\label{eq:alpha}
\end{equation}
where $\Hat{u}_p$ is a characteristic velocity scale of the turbulence intensity in the vicinity of $y_p$, defined as
\begin{equation}
\Hat{u}_p=C_\mu^{1/4}k_p^{1/2},
\end{equation}
with $C_\mu=0.09$. $\mean{U}_{p}$, $\mean{v^{\scriptscriptstyle 2}}_p$ and $k_p$ are, respectively, the mean streamwise velocity, the wall-normal component of the Reynolds stress tensor and the turbulent kinetic energy at $y_p$. In \Eqre{eq:alpha} $U_e$ is the magnitude of the equilibrium value of the mean velocity at $y_p$ and is specified by the logarithmic law
\begin{equation}
U_e = \frac{u_*}{\kappa}\log\left(E\frac{y_pu_*}{\nu}\right),
\end{equation}
where $\kappa=0.41$ is the K\'arm\'an constant and the surface roughness parameter $E=8.5$ for a smooth wall. The friction velocity $u_*$ is computed from local statistics as
\begin{equation}
u_* = \sqrt{\Hat{u}_p^2 + \gamma_\tau\left|\frac{y_p}{\rho}\frac{\partial\mean{P}}{\partial x}\right|},\label{eq:friction-velocity}
\end{equation}
with
\begin{equation}
\gamma_\tau=\max\left[0; \enskip \mathrm{sign}\left(\mean{uv}\frac{\partial\mean{P}}{\partial x}\right)\right].
\end{equation}
In Equations (\ref{eq:wall-normal-velocity}-\ref{eq:friction-velocity}) the streamwise $x$ and wall-normal $y$ coordinate directions are defined according to the local tangential and normal coordinate directions of the particular wall element in question. In other words, if the wall is not aligned with the flow coordinate system then the vectors $\mathcal{U}_i$ and $\partial\mean{P}/\partial x_i$, and the Reynolds stress tensor $\mean{u_iu_j}$, need to be appropriately transformed into the wall-element coordinate system before being employed in the above equations. 

The condition on the turbulent frequency is given by
\begin{equation}
\omega_{\scriptscriptstyle R} = \omega_{\scriptscriptstyle I}\exp\left(\beta\frac{\mathcal{V}_I}{y_p\mean{\omega}}\right),
\end{equation}
with
\begin{equation}
\beta = -\frac{2}{\tfrac{1}{2}+\tfrac{3}{4}C_0+C_3+C_{\omega2}-C_{\omega1}},\label{eq:wall-omega}
\end{equation}
which completes the description of the wall function approach.

In summary, the flow is represented by a large number of Lagrangian particles whose position $\mathcal{X}_i$, velocity $\mathcal{U}_i$, scalar concentration $\psi$ and characteristic turbulent frequency $\omega$ are governed by \Eqres{eq:Lagrangian-position}, \Eqrs{eq:Lagrangian-velocity}, \Eqrs{eq:IECM} and \Eqrs{eq:frequency-model}, respectively. These equations are discretised and advanced in time by the explicit forward Euler-Maruyama method \citep{Kloeden_99}. The mean pressure, required in \Eqre{eq:Lagrangian-velocity}, is obtained via a pressure projection scheme \citep{Bakosi_08}.  Full wall resolution is obtained through Equations (\ref{eq:Gij}-\ref{eq:L}), while wall functions are applied through Equations (\ref{eq:wall-GandC}-\ref{eq:wall-omega}). The pressure-Poisson and elliptic relaxation \Eqrs{eq:elliptic-relaxation-Lagrangian} equations are solved using an unstructured Eulerian grid with the finite element method. The grid is also used to track particles throughout the domain and to estimate Eulerian statistics using ensemble averaging. In practical simulations using PDF methods a few hundred particles per element is usually employed. Adequate stability can already be achieved using as little as 50--100 particles, however, 300--500 particles per elements are recommended to exploit the bin structure to compute $\mean{\phi|\bv{V}}$ \citep{Bakosi_08} and to decrease the statistical error. The numerical algorithm and performance issues are detailed in \citet{Bakosi_08}.

\section{Modelling the street canyon}
\label{sec:street_canyon}
\begin{figure}[t]
\centering
\resizebox{13.7cm}{!}{\input{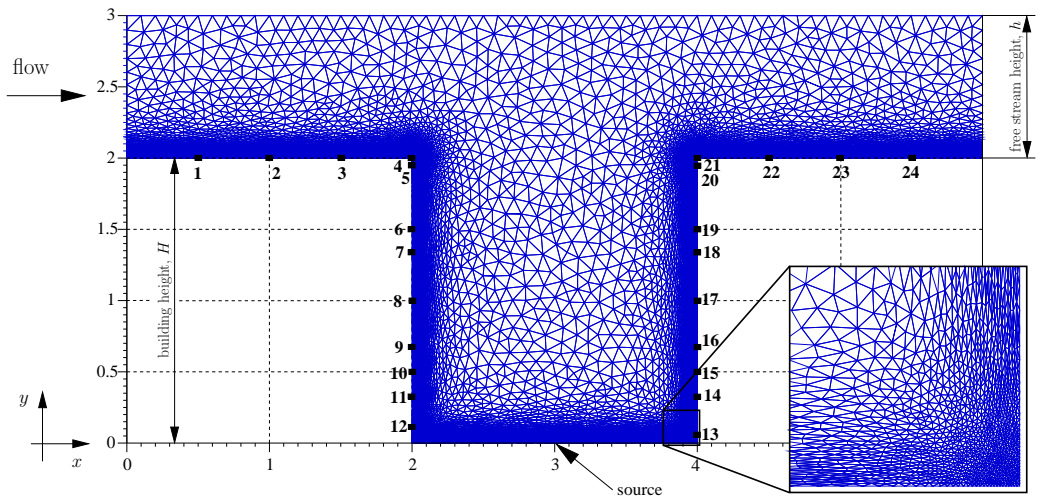}}
\caption{Geometry and Eulerian mesh (consisting of approximately 12,000 cells) for the computation of turbulent street canyon with full resolution of the wall boundary layers using elliptic relaxation. The grid is generated by the general purpose mesh generator Gmsh \citep{Geuzaine_09}. The positions labelled by bold numbers indicate the sampling locations for the passive scalar, equivalent with the combined set of measurement tapping holes of \citet{Meroney_96}, \citet{Pavageau_96} and \citet{Pavageau_99}. In the zoomed area the refinement is depicted, which ensures an adequate resolution of the boundary layer and the vortices forming in the corner.}
\label{fig:canyon-geometry}
\end{figure}
\begin{table}[b]
\caption{\label{tab:measurement-locations}Concentration sampling locations at building walls and tops according to the experimental measurement holes of \citet{Meroney_96}, \citet{Pavageau_99} and \citet{Pavageau_96}. See also \Fige{fig:canyon-geometry}.}
\begin{tabular*}{1\textwidth}{r@{\hspace{0.045\textwidth}}c@{\hspace{0.045\textwidth}}c@{\hspace{0.045\textwidth}}c@{\hspace{0.045\textwidth}}c@{\hspace{0.045\textwidth}}c@{\hspace{0.045\textwidth}}c@{\hspace{0.045\textwidth}}c@{\hspace{0.045\textwidth}}c@{\hspace{0.045\textwidth}}c@{\hspace{0.045\textwidth}}c@{\hspace{0.045\textwidth}}c@{\hspace{0.045\textwidth}}c}
\hline
\mbox{\textbf{\#} }&\textbf{1}&\textbf{2}&\textbf{3}&\textbf{4}&\textbf{5}&\textbf{6}&\textbf{7}&\textbf{8}&\textbf{9}&\textbf{10}&\textbf{11}&\textbf{12}\\
\mbox{$x$ } & \mbox{0.5 } & \mbox{1 } & \mbox{1.5 } & \mbox{2 } & \mbox{2 } & \mbox{2 } & \mbox{2 } & \mbox{2 } & \mbox{2 } & \mbox{2.5 } & \mbox{2 } & \mbox{2 } \\
\mbox{$y$ } & \mbox{2 } & \mbox{2 } & \mbox{2 } & \mbox{2 } & \mbox{1.93 } & \mbox{1.5 } & \mbox{1.33 } & \mbox{1 } & \mbox{0.67 } & \mbox{0.5 } & \mbox{0.33 } & \mbox{0.17 }\\
\hline
\mbox{\textbf{\#} }&\textbf{13}&\textbf{14}&\textbf{15}&\textbf{16}&\textbf{17}&\textbf{18}&\textbf{19}&\textbf{20}&\textbf{21}&\textbf{22}&\textbf{23}&\textbf{24}\\
\mbox{$x$ } & \mbox{4 } & \mbox{4 } & \mbox{4 } & \mbox{4 } & \mbox{4 } & \mbox{4 } & \mbox{4 } & \mbox{4 } & \mbox{4 } & \mbox{4.5 } & \mbox{5 } & \mbox{5.5 }\\
\mbox{$y$ } & \mbox{0.17 } & \mbox{0.33 } &\mbox{0.5 } & \mbox{0.67 } & \mbox{1 } & \mbox{1.33 } & \mbox{1.5 } & \mbox{1.93 } & \mbox{2 } & \mbox{ 2 } & \mbox{2 } & \mbox{ 2 }\\
\hline
\end{tabular*}
\end{table}
\begin{table}
\caption{\label{tab:constants}Constants for modelling the joint PDF of velocity, characteristic turbulent frequency and transported passive scalar.}
\begin{tabular*}{1\textwidth}{c@{\hspace{0.05\textwidth}}c@{\hspace{0.05\textwidth}}c@{\hspace{0.05\textwidth}}c@{\hspace{0.05\textwidth}}c@{\hspace{0.05\textwidth}}c@{\hspace{0.05\textwidth}}c@{\hspace{0.05\textwidth}}c@{\hspace{0.05\textwidth}}c@{\hspace{0.05\textwidth}}c@{\hspace{0.05\textwidth}}c@{\hspace{0.05\textwidth}}c@{\hspace{0.05\textwidth}}c@{\hspace{0.05\textwidth}}}
\hline
$C_1$&$C_2$&$C_3$&$C_4$&$C_T$&$C_L$&$C_\eta$&$C_v$&$\gamma_5$&$C_{\omega1}$&$C_{\omega2}$&$C_s$&$C_t$\\
\hline
1.85&0.63&5&0.25&6&0.134&72&1.4&0.1&0.5&0.73&0.02&0.7\\
\hline
\end{tabular*}
\end{table}
\begin{figure}[t]
\centering
\includegraphics[width=13.7cm]{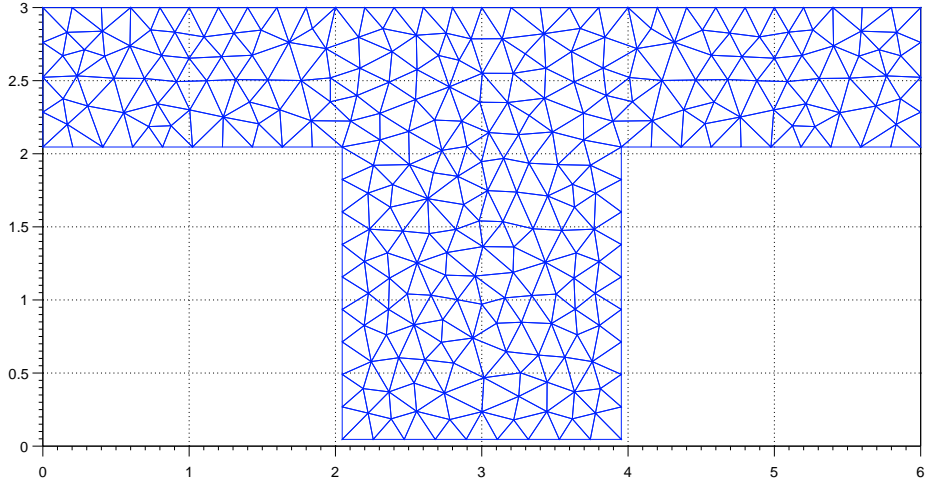}
\caption{Geometry and Eulerian mesh (consisting of approximately 500 cells) for the computation of flow in a turbulent street canyon with wall functions at $\textit{Re}\approx12000$. The domain is stripped at no-slip walls so that it does not include the close vicinity of the wall at $y^+<30$. The positions for sampling the scalar concentrations are the same as in \Fige{fig:canyon-geometry}.}
\label{fig:canyon-wf-geometry}
\end{figure}
Street canyons are often used to study flow and pollutant dispersion patterns in urban areas. A wealth of experimental data for this simplified urban-scale setting is available from wind-tunnel and LES data \citep{Meroney_96,Pavageau_99,Liu_02}, making it a natural choice to validate the current, newly developed method. We will simulate the ``urban roughness'' case of \citet{Meroney_96}, which is a model for a series of street canyons in the streamwise direction. The simulations are performed for statistically two-dimensional flow geometry, with periodic inflow and outflow boundary conditions in the free stream above the buildings (i.e.\ the particles crossing the outflow boundary are re-injected at the inflow boundary). The Reynolds number based on the maximum free stream velocity $U_0$ and the building height $H$ was $\textit{Re}\approx12000$. This corresponds to $\textit{Re}_\tau\approx600$ based on the friction velocity and the free stream height, $h=H/2$, if the free stream above the buildings is considered as the lower part of an approximate fully developed turbulent channel flow. The velocity-conditioned scalar mean $\mean{\phi|\bv{V}}$ required in \Eqre{eq:IECM} has been computed using the general method described by \citet{Bakosi_07} using a bin structure of $(5\times5\times5)$. After the flow has reached a statistically stationary state, time averaging is used to collect velocity statistics and a continuous scalar is released from a street level line source at the centre of the canyon (corresponding to a point source in two dimensions). Particles travelling through the source of diameter $0.01/h$ are assigned a unit source strength. The scalar field is also time averaged after it has reached a stationary state. 

The simulations with the full resolution model have been made with the constants given in Table \ref{tab:constants}, using 300 particles per element. The Eulerian mesh used for this simulation is displayed in \Fige{fig:canyon-geometry}, which shows the considerable refinement along the building walls and tops necessary to resolve the boundary layers. In this case, the high anisotropy and inhomogeneity of the Reynolds stress tensor in the vicinity of walls are captured by the elliptic relaxation technique, using \Eqre{eq:elliptic-relaxation-Lagrangian}. 

The simulations using wall functions were performed on the Eulerian mesh displayed in \Fige{fig:canyon-wf-geometry}, also using 300 particles per element. The particle-boundary conditions described in Sec. \ref{sec:wall-functions} were implemented for arbitrary geometry. Note that the first grid point where the boundary conditions based on wall functions are to be applied should not be closer to the wall than $y^+=y u_\tau/\nu=30$, where $y^+$ is the non-dimensional distance from the wall in wall units, but should still be sufficiently close to the wall to lie in the inertial sublayer \citep{Dreeben_97b}. Accordingly, the grid in \Fige{fig:canyon-wf-geometry} only contains the domain stripped from the wall region at $y^+<30$.

Turbulence and scalar statistics are obtained entirely from the particles that represent both the flow itself and the scalar concentration field. The Eulerian meshes displayed in \Fige{fig:canyon-geometry} for the full resolution and in \Fige{fig:canyon-wf-geometry} for the wall function cases are used to extract the statistics, to track the particles throughout the domain and to solve the Eulerian equations, namely \Eqre{eq:elliptic-relaxation-Lagrangian} and the mean-pressure-Poisson equation in the fully resolved case and only the latter in the wall function case.
\begin{figure}
\centering
\psfragscanon
\includegraphics[width=8.1cm]{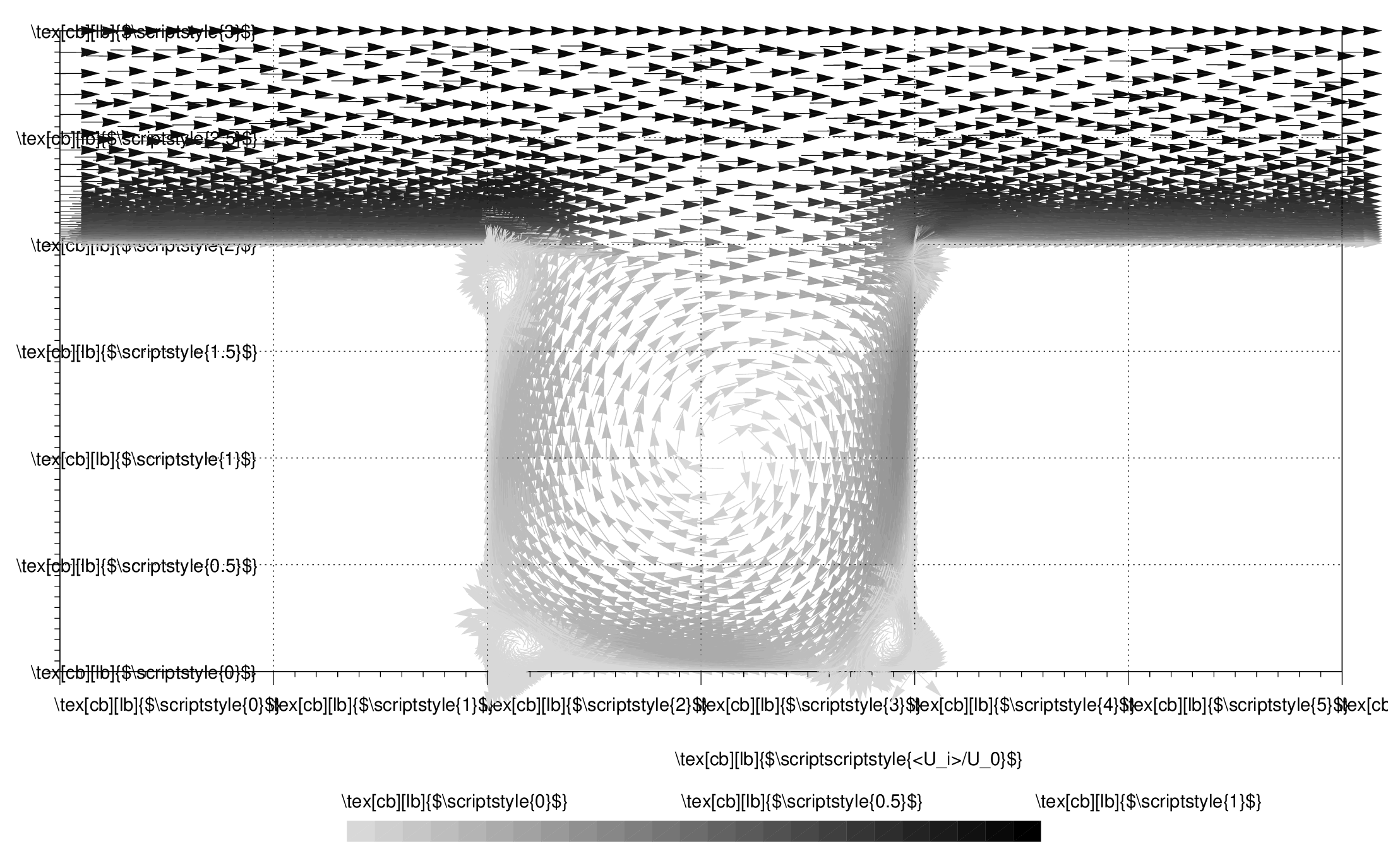}
\includegraphics[width=8.1cm]{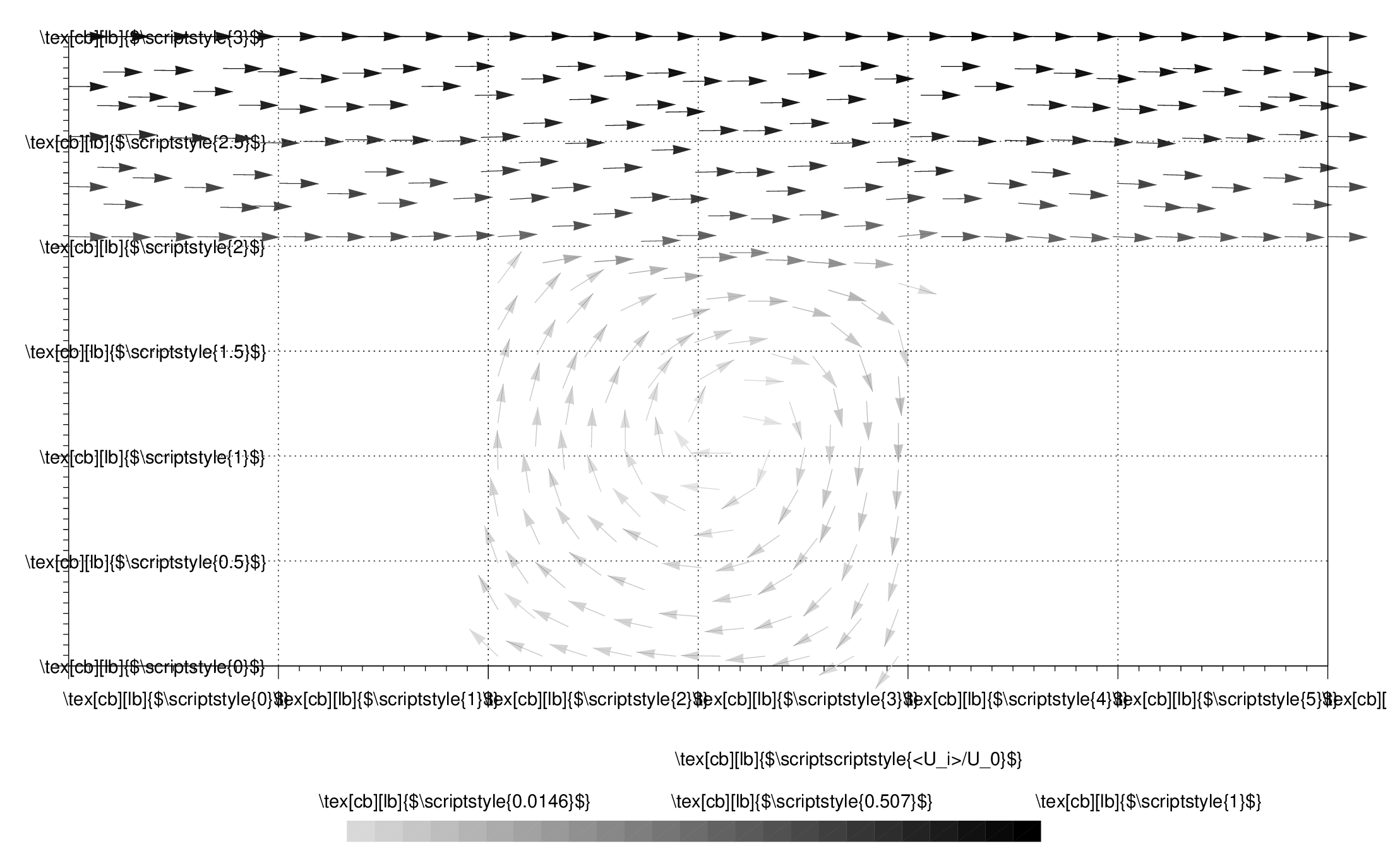}
\includegraphics[width=8cm]{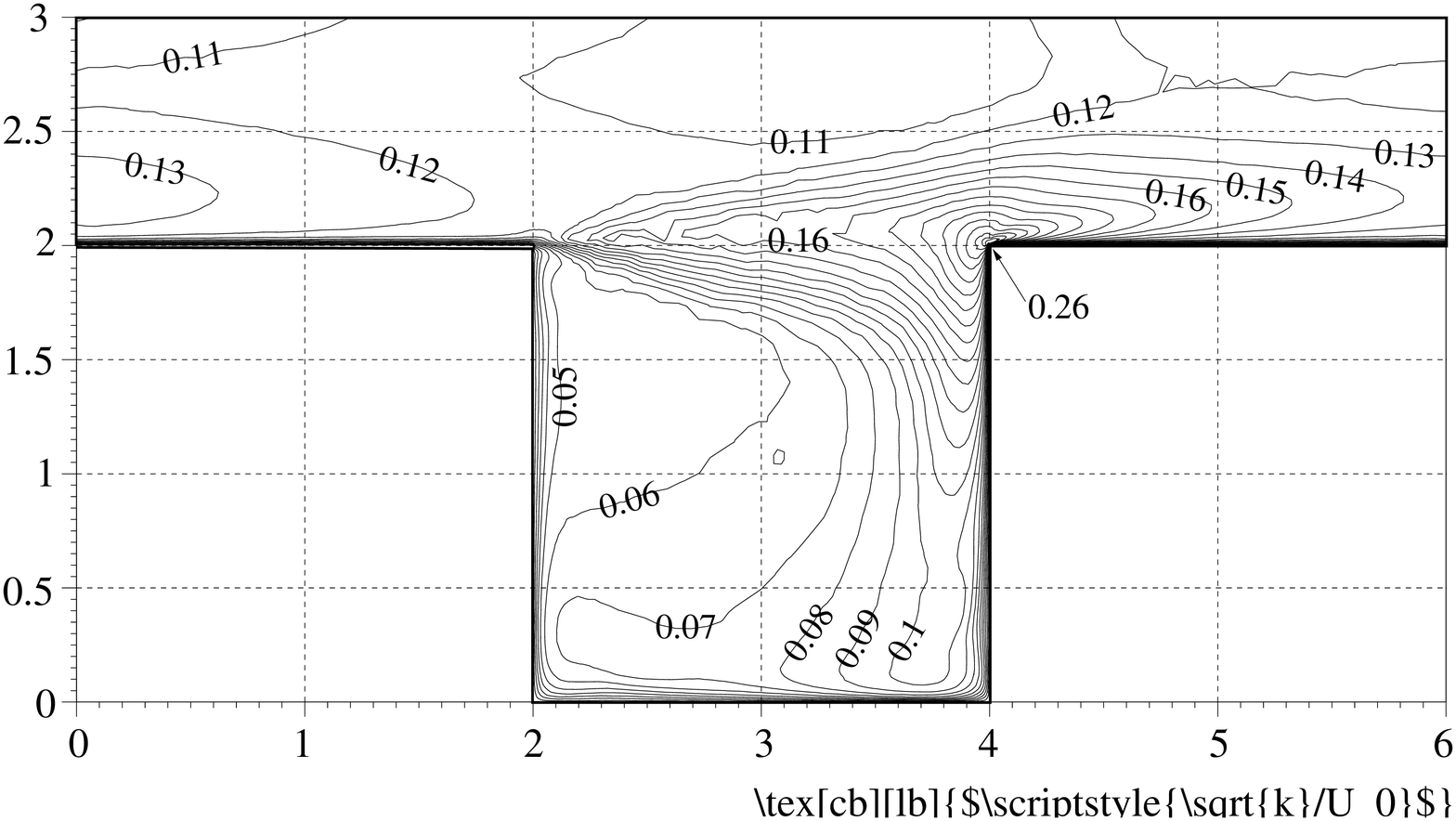}
\includegraphics[width=8cm]{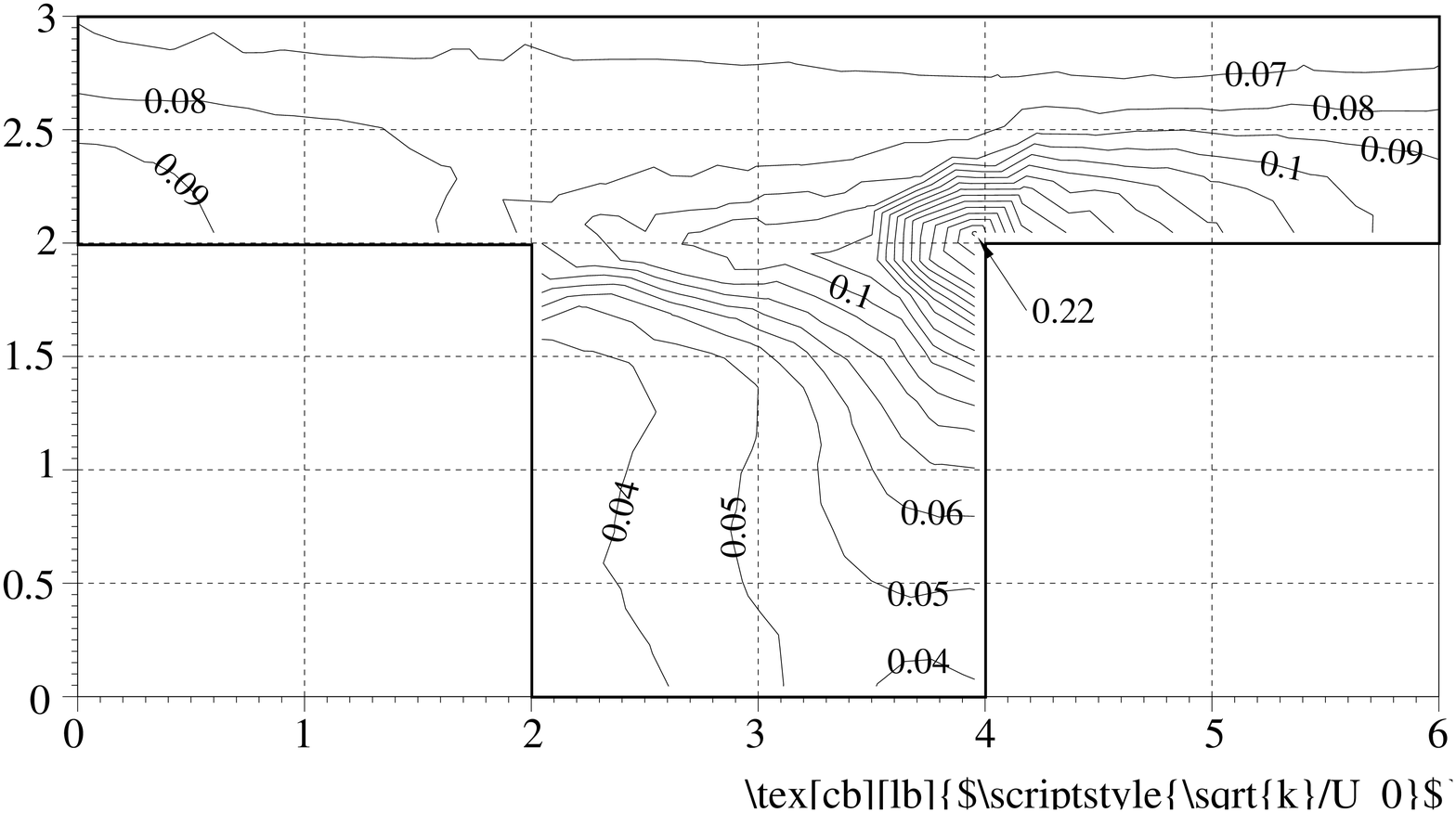}
\psfragscanoff
\caption{Velocity vectors (first row) and iso-contours of turbulent kinetic energy (second row) of the fully developed turbulent street canyon at $\textit{Re}\approx12000$ based on the maximum free stream velocity $U_0$ and the building height $H$. Left -- full resolution with elliptic relaxation, right -- coarse simulation with wall functions.}
\label{fig:canyon-velocity}
\end{figure}

In \Fige{fig:canyon-velocity}, the mean velocity vector field and the iso-contours of the turbulent kinetic energy are displayed for both fully resolved and wall function simulations.  It is apparent that the full resolution captures even the smaller counter-rotating eddies at the internal corners of the canyon, while the coarse grid resolution with wall functions only captures the overall flow pattern characteristic of the flow, such as the large steadily rotating eddy inside the canyon. The turbulent kinetic energy field is captured in a similar manner.  Both methods reproduce the highest turbulence activity at the building height above the canyon, with a maximum at the windward building corner.  The full resolution simulation shows a more detailed spatial distribution of energy, whereas the coarse resolution of the wall-function simulation still allows tha capture of the overall pattern. 

\begin{figure}
\centering
\includegraphics[width=8cm]{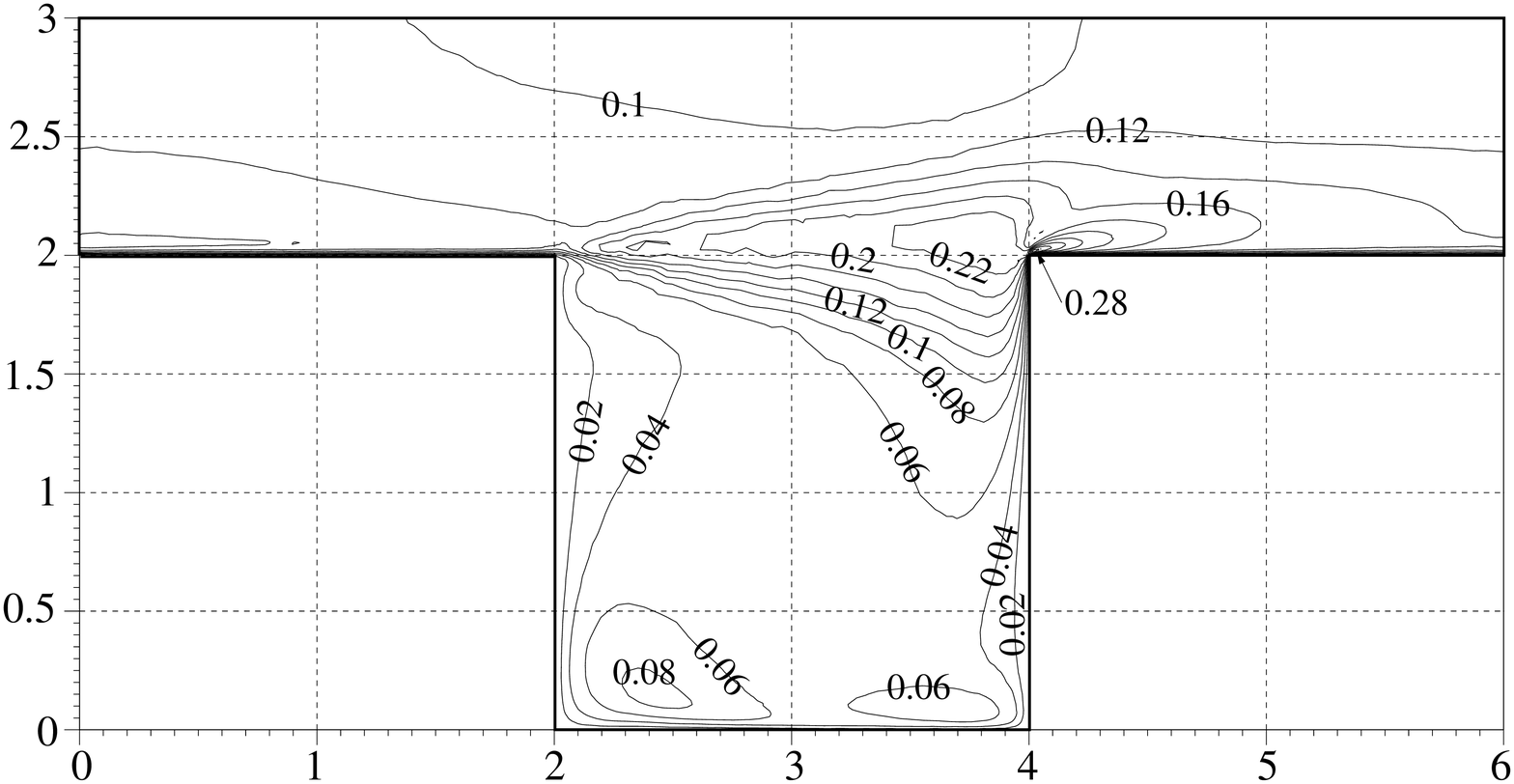}
\includegraphics[width=8cm]{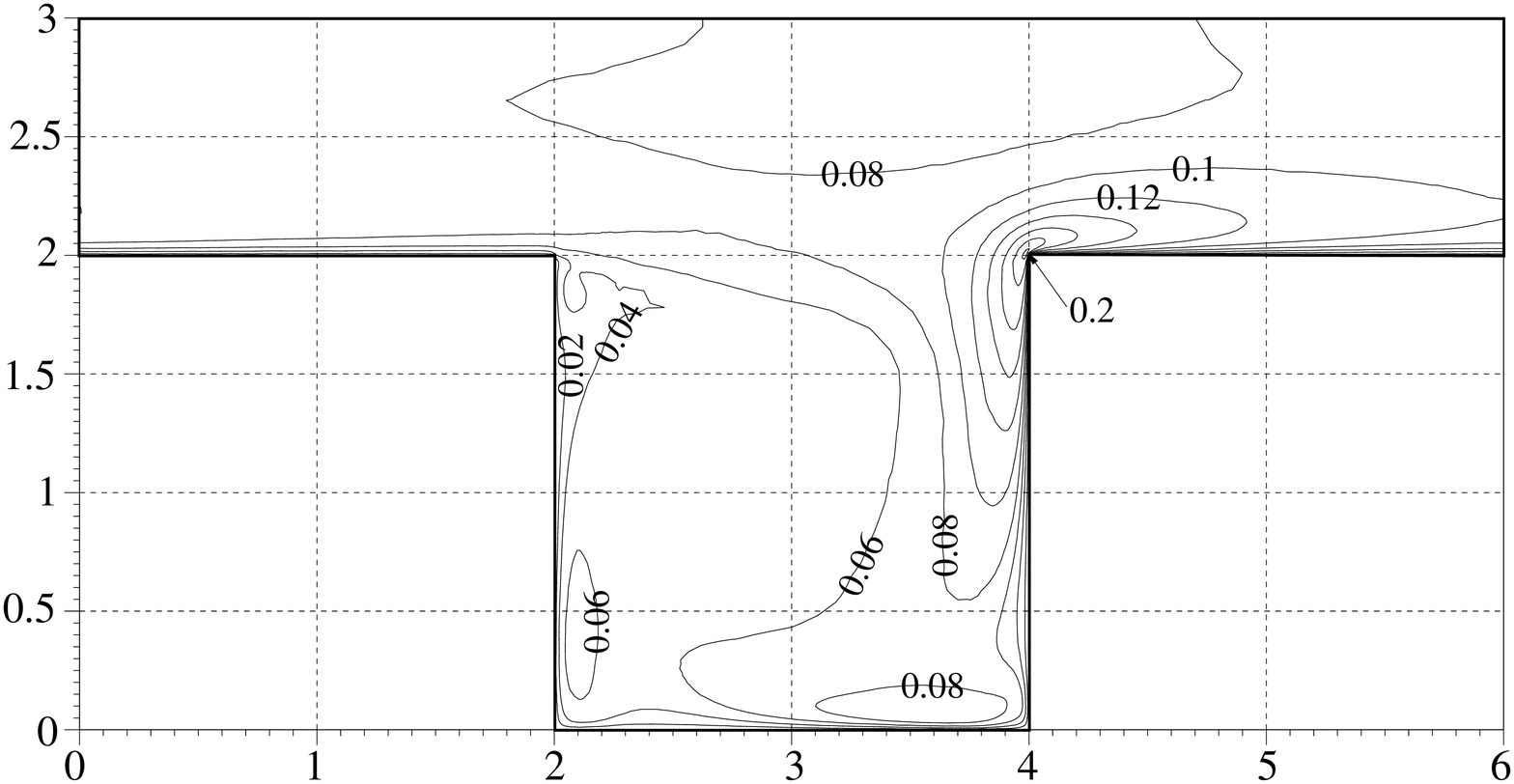}
\includegraphics[width=8cm]{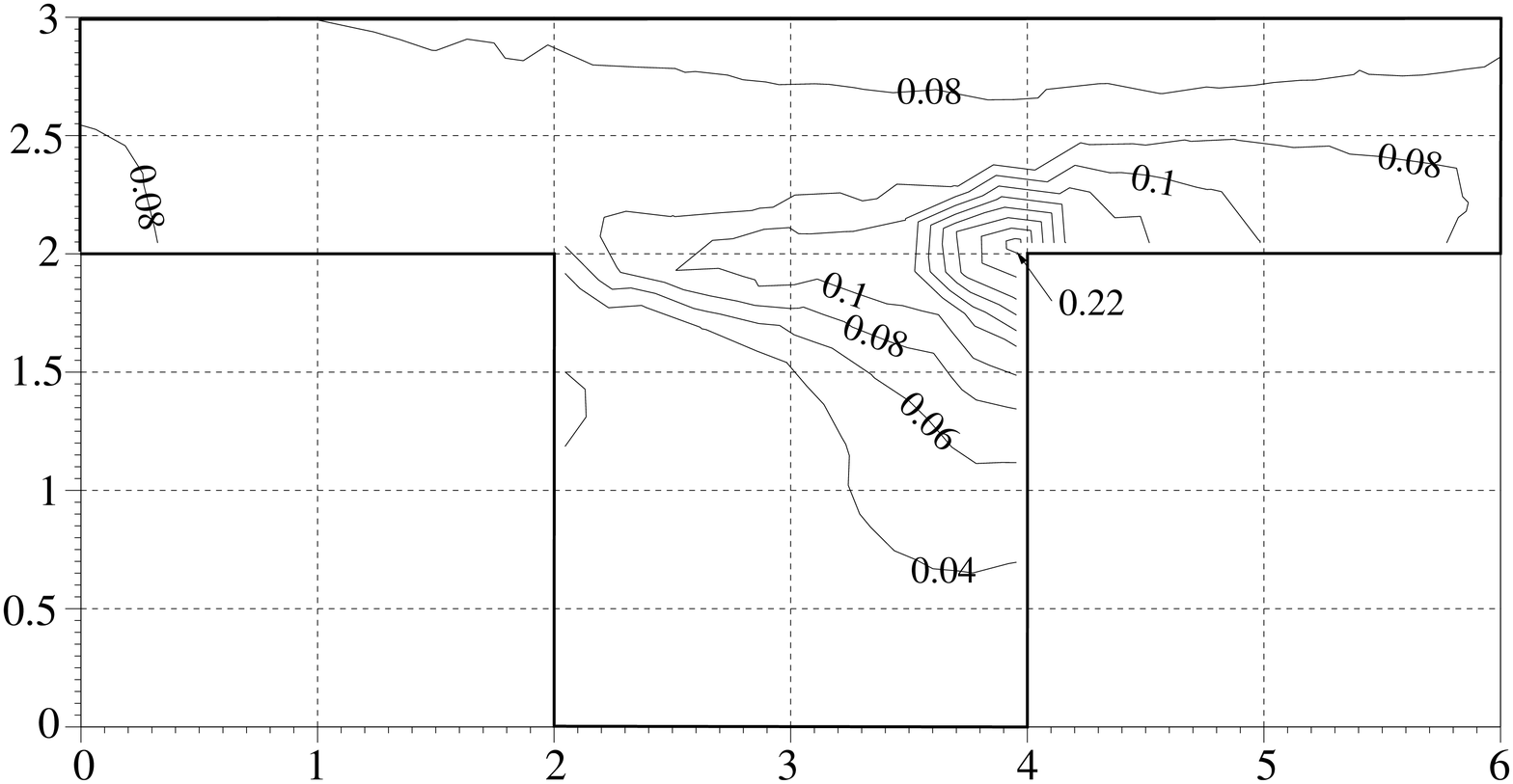}
\includegraphics[width=8cm]{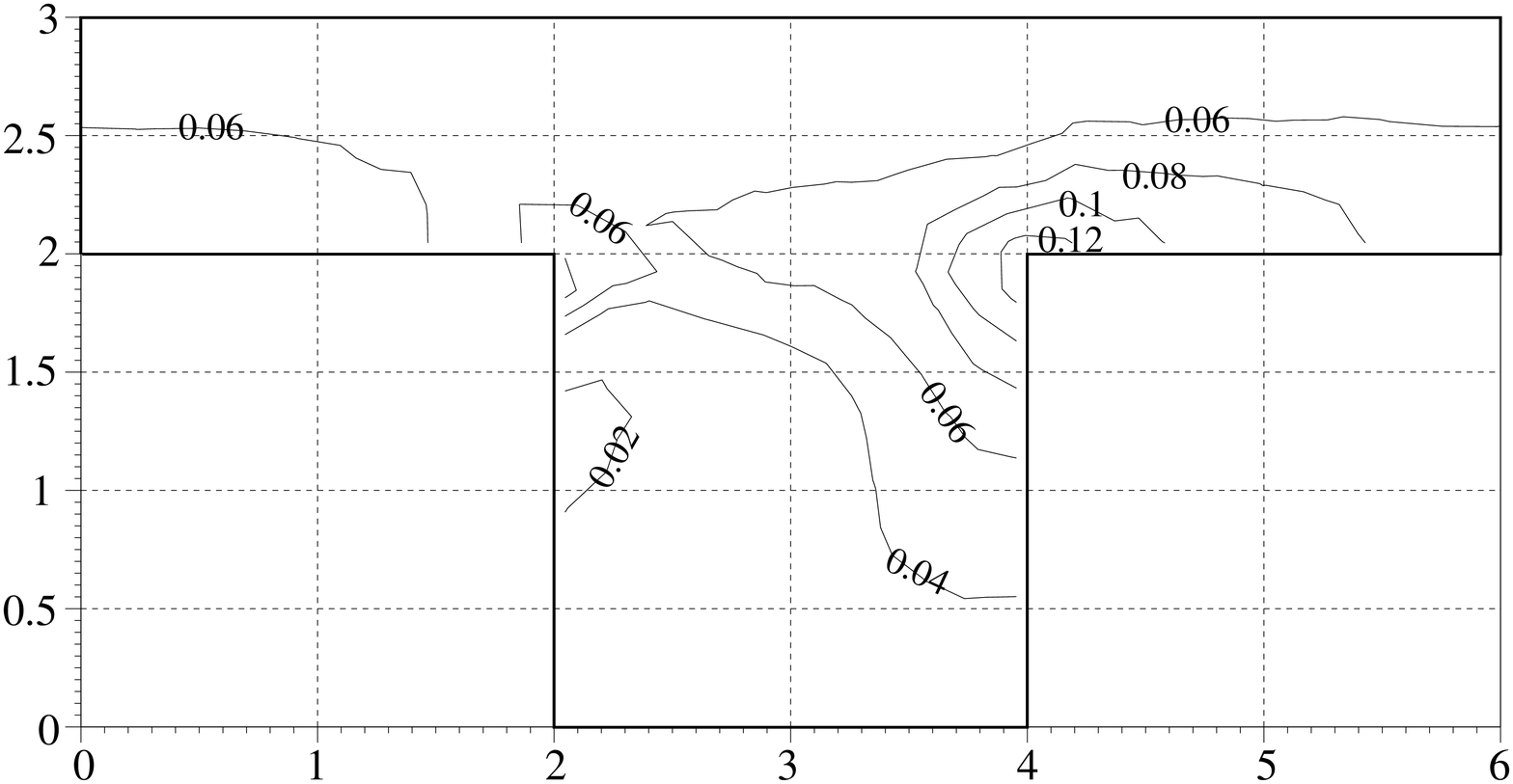}
\includegraphics[width=8cm]{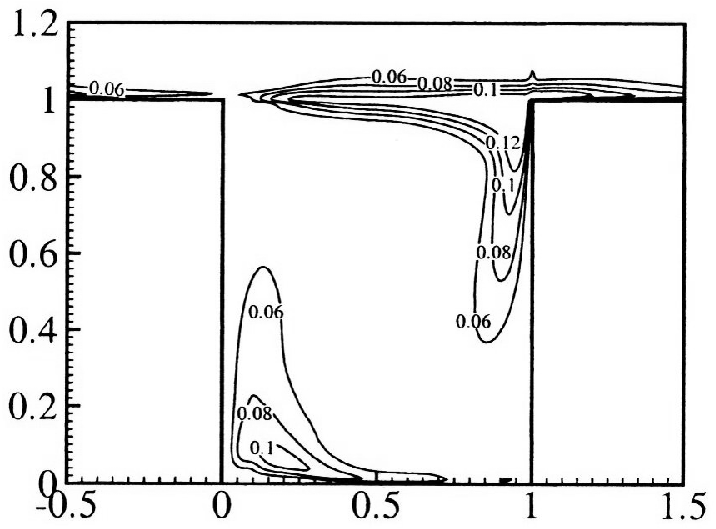}
\includegraphics[width=8cm]{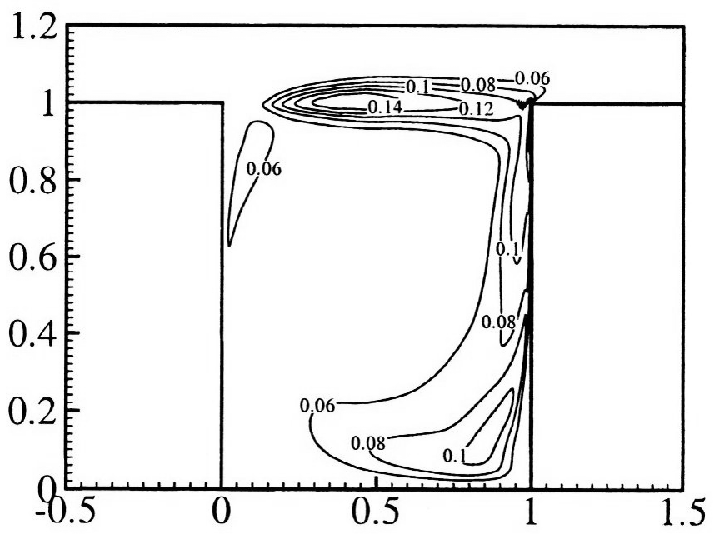}
\caption{Dimensionless turbulent intensities $\mean{u^2}^{\scriptscriptstyle 1/2}/U_0$ (first column) and $\mean{w^2}^{\scriptscriptstyle 1/2}/U_0$ (second column) computed using full wall resolution (first row) and using wall functions (second row) at $\textit{Re}\approx12000$ compared with the LES results (third row) of \citet{Liu_02}.}
\label{fig:canyon-Reynolds}
\end{figure}
In \Fige{fig:canyon-Reynolds} the normalized turbulent intensities $\mean{u^{\scriptscriptstyle 2}}^{\scriptscriptstyle 1/2}/U_0$ and $\mean{w^{\scriptscriptstyle 2}}^{\scriptscriptstyle 1/2}/U_0$ are displayed for both simulation cases and compared with the LES results of \citet{Liu_02}. In the large eddy simulations the filtered momentum equations are solved by the Galerkin finite element method using brick three-dimensional elements, while the residual stresses are modelled using the Smagorinsky closure.

The full resolution simulation shows very good agreement with LES. The contour plots of $\mean{u^{\scriptscriptstyle 2}}^{\scriptscriptstyle 1/2}/U_0$ correctly display two local maxima, at the windward external and at the leeward internal corners. The contour plots of $\mean{w^{\scriptscriptstyle 2}}^{\scriptscriptstyle 1/2}/U_0$ show distributed high values at the building level above the canyon, along the windward internal corner and wall, and at the street level downstream of the source. By contrast, the wall-function contour plots are in general less detailed, failing to reproduce the internal maximum of $\mean{u^{\scriptscriptstyle 2}}^{\scriptscriptstyle 1/2}/U_0$, and showing a more uniform representation of $\mean{w^{\scriptscriptstyle 2}}^{\scriptscriptstyle 1/2}/U_0$.

Several wind-tunnel measurements have been carried out for a scalar released from a street level continuous line source at the centre of the canyon, providing concentration statistics above the buildings, at the walls, and inside the canyon \citep{Meroney_96,Pavageau_96,Pavageau_99}. To examine the concentration values along the building walls and tops, we sampled the computed mean concentration field at the locations depicted in \Fige{fig:canyon-geometry} and listed in Table \ref{tab:measurement-locations}.

\begin{figure}
\centering
\psfragscanon
\rotatebox{270}{\includegraphics[width=8cm]{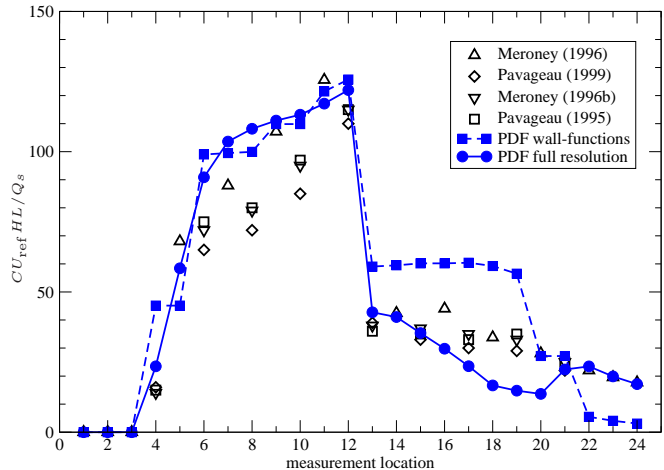}}
\psfragscanoff
\caption{Distribution of mean concentrations at the boundary of the street canyon. The experimental data are in terms of the ratio $CU_\mathrm{ref}HL/Q_\mathrm{s}$, where $C$ is the actual measured mean concentration (ppm), $U_\mathrm{ref}$ is the free-stream mean velocity ($\mathrm{m}\mathrm{s}^{-1}$) taken at the reference height $y_\mathrm{ref}\approx11H$ and $Q_\mathrm{s}/L$ is the line source strength ($\mathrm{m}^2\mathrm{s}^{-1}$) in which $Q_\mathrm{s}$ denotes the scalar flow rate and $L$ is the source length. The calculation results are scaled to the concentration range of the experiments. References for experimental data: $\scriptscriptstyle{\triangle}$ \citet{Meroney_96}; $\diamond$, $\scriptscriptstyle\triangledown$, \citet{Pavageau_99}; $\scriptscriptstyle\square$ \citet{Pavageau_96}. See also \Fige{fig:canyon-geometry} and Table \ref{tab:measurement-locations} for the measurement locations.}
\label{fig:wall-concentrations}
\end{figure}

The excellent agreement of the results using both full resolution and wall functions with a number of experiments is shown in \Fige{fig:wall-concentrations}. The concentration peak is precisely captured at the internal leeward corner and the model accurately reproduces the pattern of concentration along both walls including the higher values along the leeward wall.
\begin{figure}
\centering
\includegraphics[width=7.5cm]{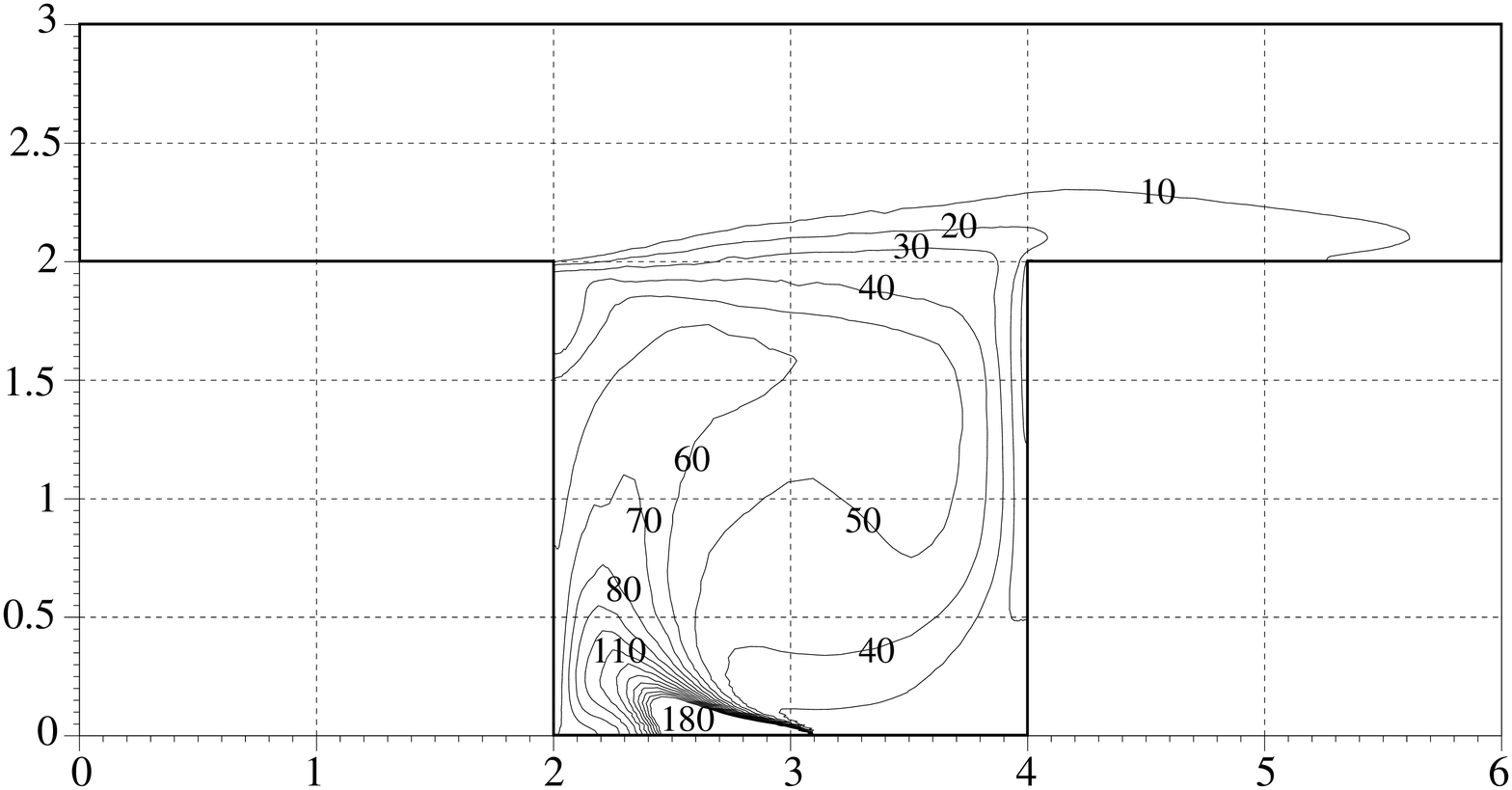}
\includegraphics[width=7.5cm]{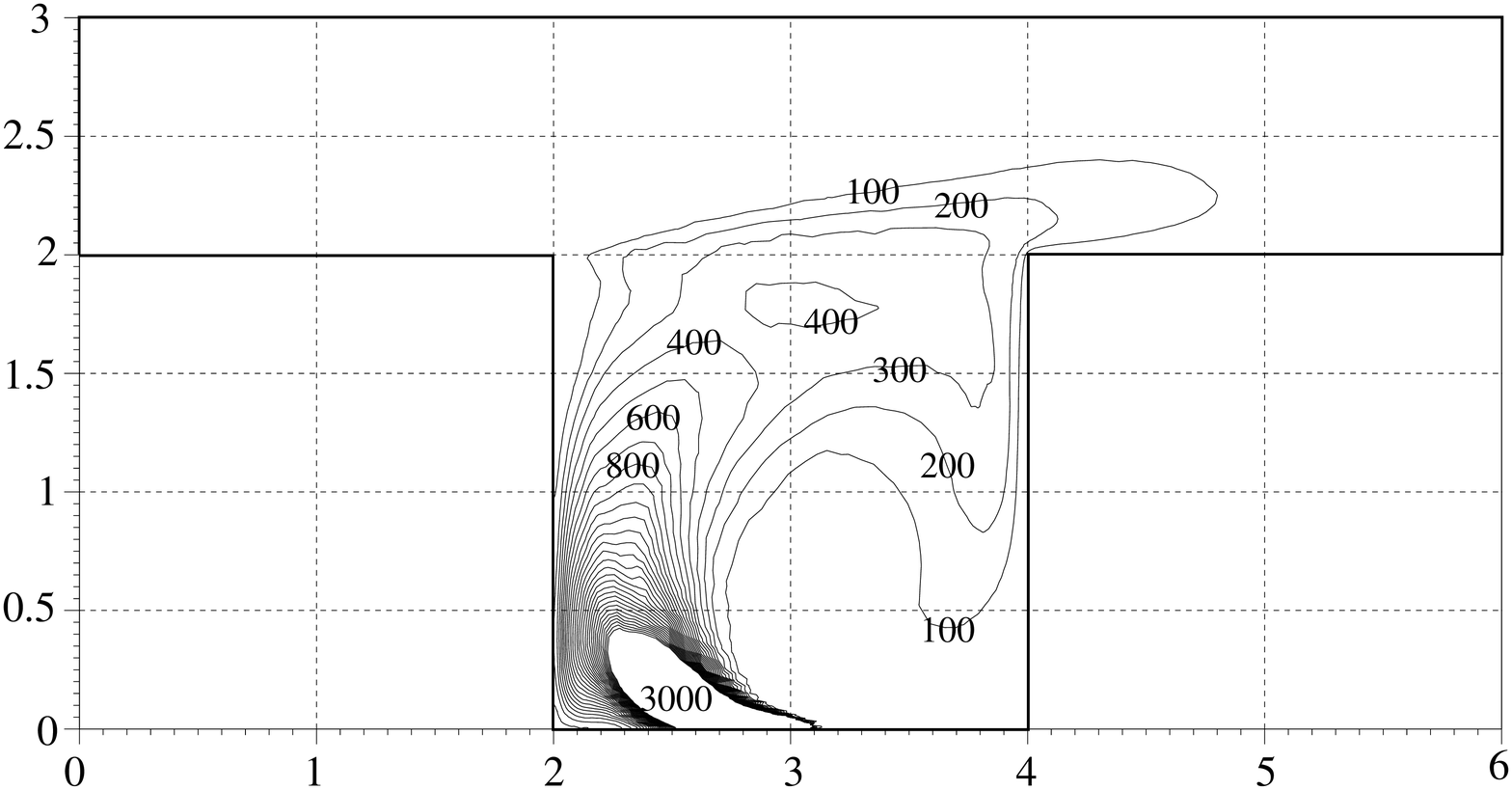}
\includegraphics[width=7.5cm]{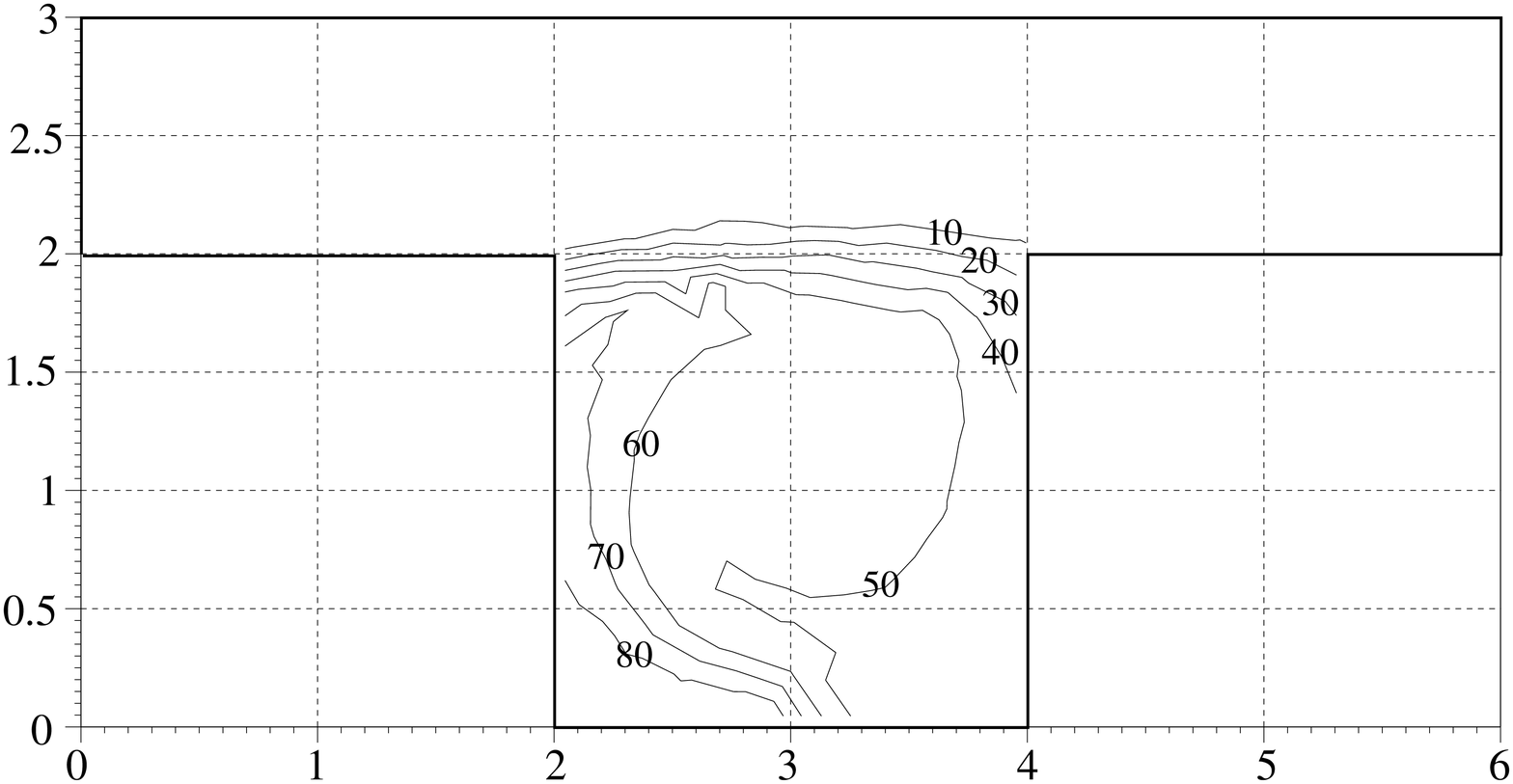}
\includegraphics[width=7.5cm]{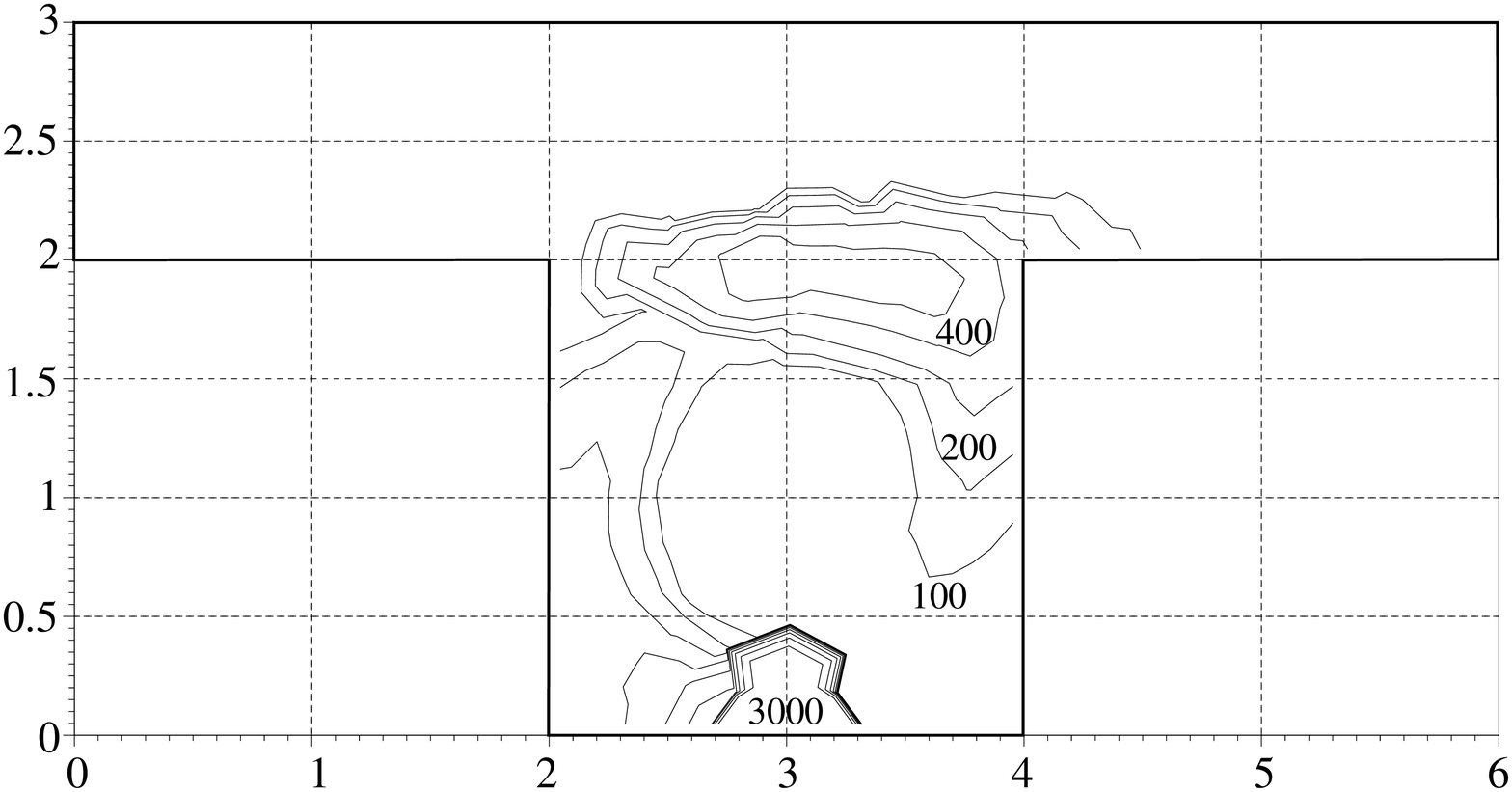}
\includegraphics[width=7.5cm]{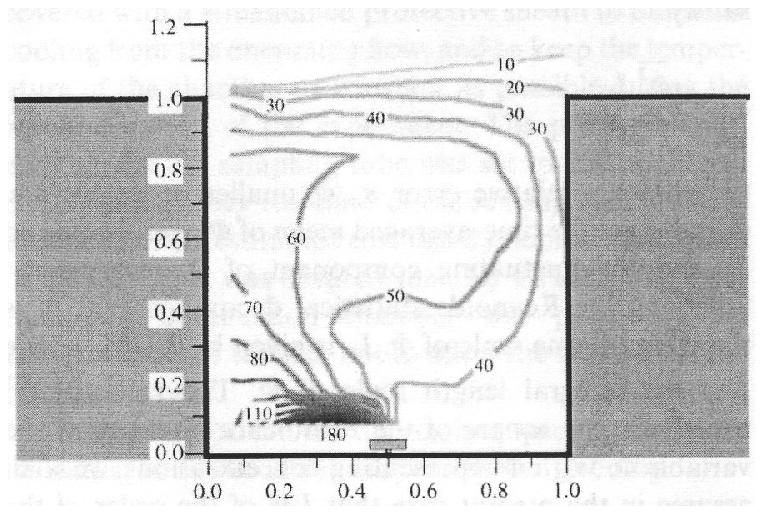}
\includegraphics[width=7.5cm]{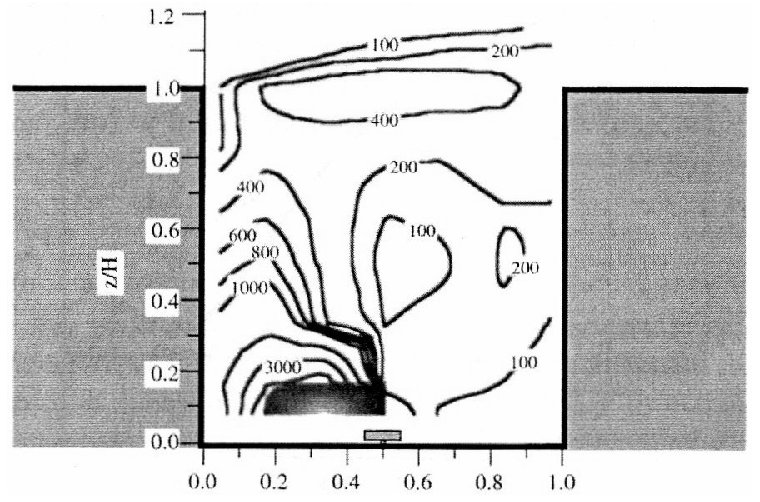}
\includegraphics[width=7.5cm]{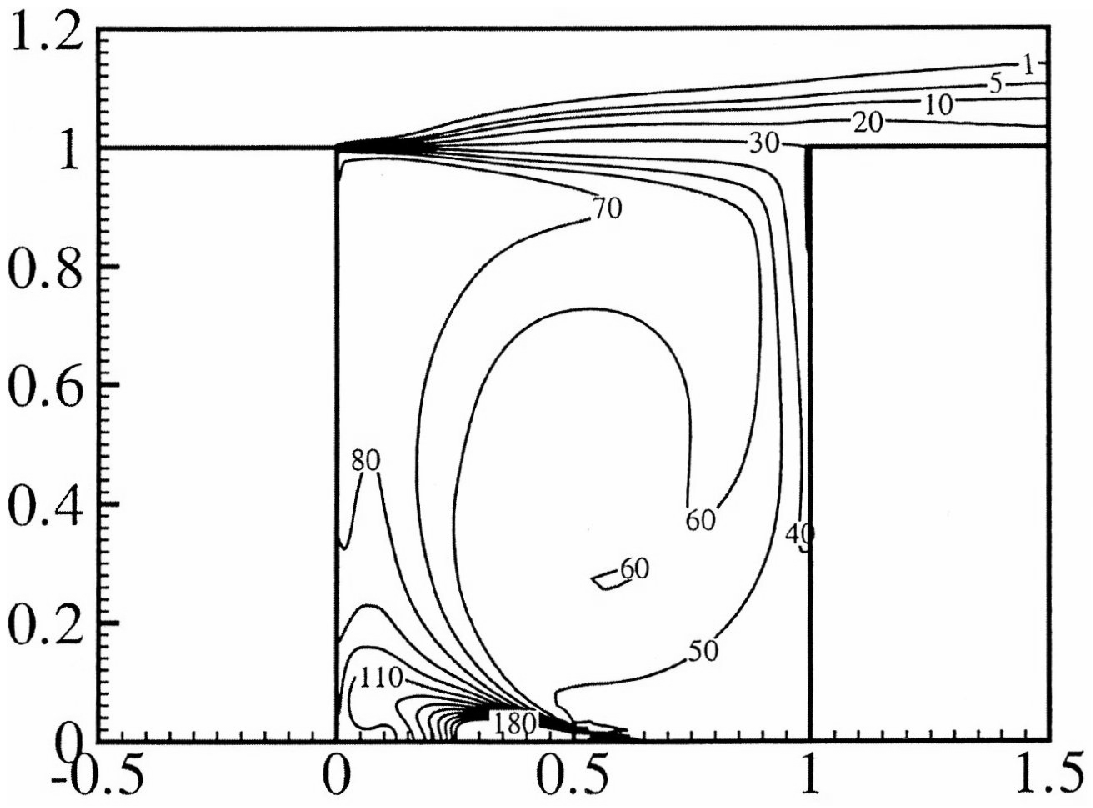}
\includegraphics[width=7.5cm]{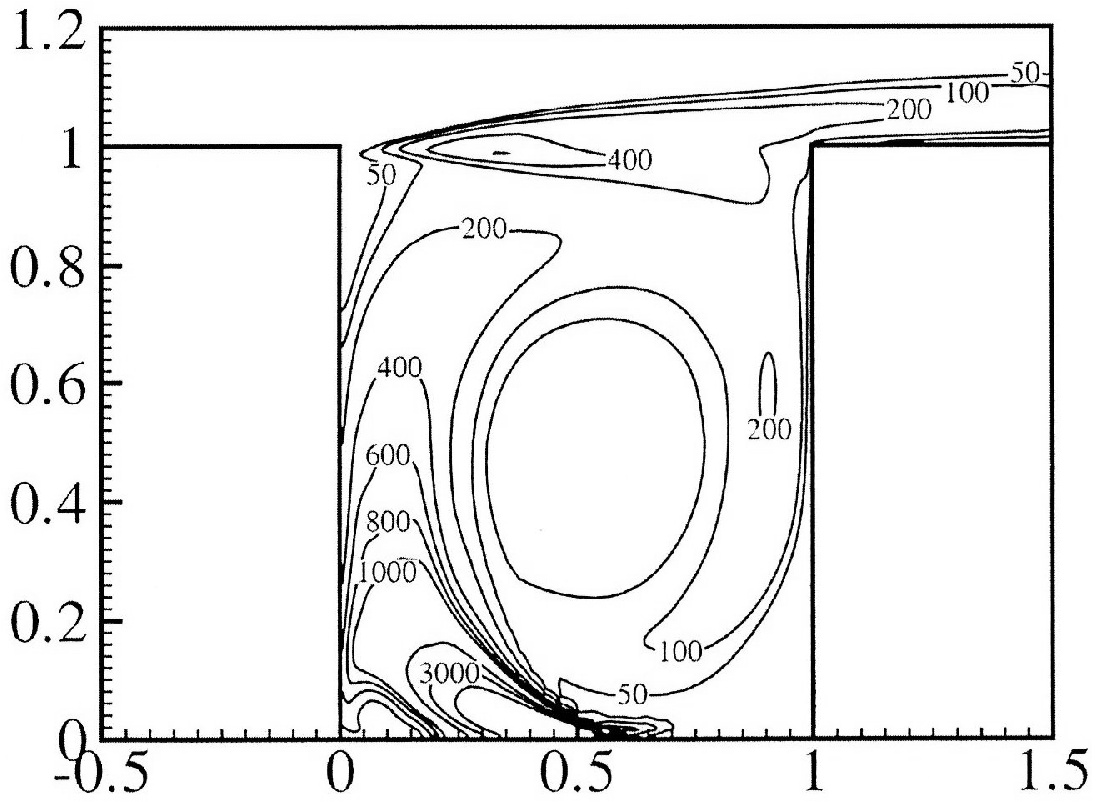}
\caption{Comparison of the spatial distribution of the normalized mean $CU_\mathrm{ref}HL/Q_\mathrm{s}$ (left column) and variance $\mean{c^2}(U_\mathrm{ref}HL/Q_\mathrm{s})^2$ (right column) of the scalar released at the centre of the street level. The normalization and the scaling of the calculated results are the same as in \Fige{fig:wall-concentrations}. First row -- PDF calculations with full wall resolution, second row -- PDF calculations with wall functions, third row -- experimental data of \citet{Pavageau_99} and fourth row -- LES calculations of \citet{Liu_02}.}
\label{fig:scalar-mean-variance}
\end{figure}

In \Fige{fig:scalar-mean-variance}, the first two statistical moments of the concentration inside the canyon are compared with experimental data and LES. The agreement with observations indicates that both the fluid dynamics and the micro-mixing components of the model provide a good representation of the real field.

\begin{figure}
\centering
\resizebox{16cm}{!}{\input{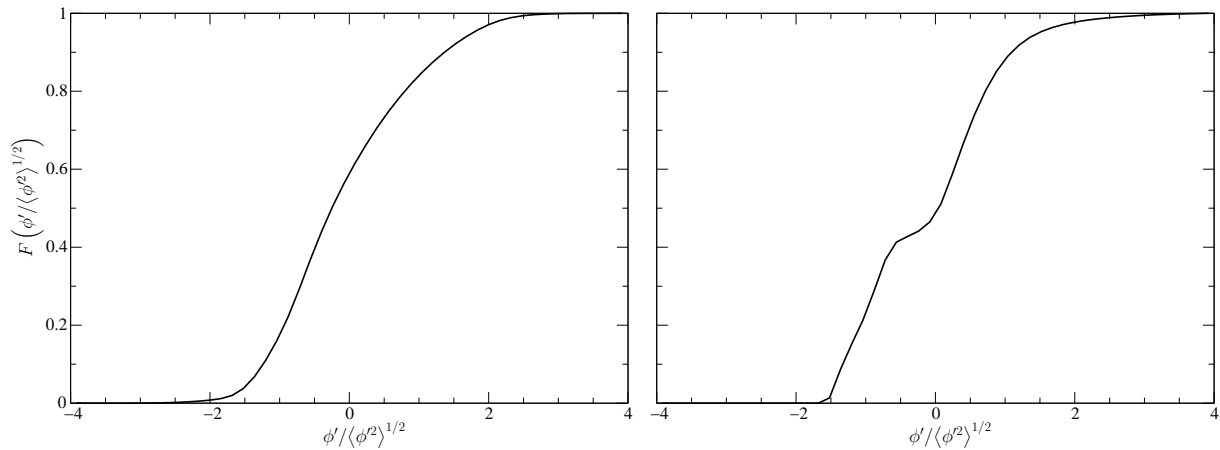}}
\caption{Cumulative distribution functions of scalar concentration fluctuations (left) at $x=3,$ $y=0.2$ and (right) at $x=3,$ $y=2$ using full resolution at walls.}
\label{fig:cdfs}
\end{figure}

Because the one-point one-time joint PDF contains all higher-order statistics and correlations of the velocity and scalar fields resulting from a close, low-level interaction between the two fields, a great wealth of statistical information is available for atmospheric transport and dispersion calculations. As an example, in \Fige{fig:cdfs} the cumulative distribution functions (CDF) of scalar concentration fluctuations
\begin{equation}
F(\phi)=\int_{\psi=0}^{\phi}\int_{\mathbf{V}=-\infty}^{\infty} f \mathrm{d}\bv{V}\mathrm{d}\psi
\end{equation}
are depicted after time averaging at selected locations of the domain for the full resolution case. No experimental data are available to assess the distributions in \Fige{fig:cdfs}, although the irregular shape of the CDF at $x=3$, $y=2$, which corresponds to a multimodal PDF, is likely to be an artifact of the micro-mixing model.

The performance gain obtained by applying wall functions as opposed to full resolution was about two orders of magnitude already at our moderate Reynolds number. The gain for higher Reynolds numbers is expected to increase faster than linearly.  

\section{Discussion}
\label{sec:discussion}
We have used an Eulerian unstructured grid, consisting of triangular element types, to estimate Eulerian statistics, to track particles throughout the domain, and to solve for inherently Eulerian quantities in conjunction with a PDF method. The boundary layers developing close to solid walls are fully captured with an elliptic relaxation technique, but can also be represented by wall functions, which use a coarser grid resolution and require significantly less particles, resulting in substantial savings in computational cost. We found that the one-point statistics of the joint PDF of velocity and scalar are well-captured by the wall function approximation.  In view of its affordable computational load and reasonable accuracy, this approximation appears to hold a realistic potential for application of the PDF method in atmospheric simulations, where the natural extension of the work is the implementation of the model in three spatial dimensions.

In hybrid PDF models developed for complex chemically reacting flows, numerical treatments for boundary conditions have been included for symmetric, inflow, outflow and free-slip walls employing the ghost-cell approach common in finite volume methods \citep{Rembold_06}. The representation of no-slip boundaries adds a significant challenge to the above cases. This is partly due to the increased computational expense entailed by the higher Eulerian grid resolution that is required to fully resolve the boundary layers. In addition, there is an increased complexity in specifying the no-slip particle conditions for both fully resolved and wall function representations. We presented an implementation of both approaches to treat no-slip boundaries with unstructured grids in conjunction with the finite element method. This also obviates further complications with ghost cells.

In the case of full wall resolution we employed the Lagrangian equivalent of a modified isotropisation of production (IP) model as originally suggested by \citet{Dreeben_98}. The elliptic relaxation technique, however, allows for the application of any turbulence model developed for high-Reynolds-number turbulence \citep{Durbin_93,Whizman_96}. The standard test case for developing near-wall models is the fully developed turbulent channel flow. In this case, we explored the simpler \citet{Rotta_51} model, which is the Eulerian equivalent of the simplified Langevin model (SLM) in the Lagrangian framework \citep{Pope_94}. This is simply achieved by eliminating the term involving the fourth-order tensor $H_{ijkl}$ from the right-hand side of \Eqre{eq:elliptic-relaxation-Lagrangian}. While the SLM makes no attempt to represent the effect of rapid pressure \citep{Pope_00} (in fact it is strictly correct only in decaying homogeneous turbulence), it is widely applied due to its is simplicity and robustness. Our experience showed a slight degradation of the computed velocity statistics (as compared to direct numerical simulation) using SLM for the case of channel flow. Since we experienced no significant increase in computational expense or decrease in numerical stability, we retained the original IP model.

Similarly, in the case of wall functions, several choices are available regarding the employed turbulence model. The methodology developed by \citet{Dreeben_97b} uses the SLM, but it is general enough to include other more complex closures, such as the \citet{Haworth_86,Haworth_87} models (HP1 and HP2), the different variants of the IP models (IPMa, IPMb, LIPM) \citep{Pope_94} or the Lagrangian version of the SSG model of \citet{Speziale_91}. All these closures can be collected under the umbrella of the generalized Langevin model, by specifying its constants as described by \citet{Pope_94}. These models have been all developed for high-Reynolds-number turbulence and need to be modified in the vicinity of no-slip walls. Including them in the wall-function formulation is possible by specifying the reflected particle frequency at the wall as
\begin{equation}
\omega_{\scriptscriptstyle R} = \omega_{\scriptscriptstyle I}\exp({-2\mathcal{V}_I\mean{\omega v}_{\scriptscriptstyle p}}/{\mean{\omega v^{\scriptscriptstyle 2}}_{\scriptscriptstyle p}})
\end{equation} instead of \Eqre{eq:wall-omega}. This involves the additional computation of the statistics \(\mean{\omega v}\) and \(\mean{\omega v^{\scriptscriptstyle 2}}\) at \(y_p\), which does not increase the computational cost significantly, but may result in a numerically less stable condition since the (originally constant) parameter \(\beta\) that appears using the SLM has been changed to a variable that fluctuates during simulation. We implemented and tested all the above turbulence models using the wall function technique. Without any modification of the model constants we found the IPMa and SLM to be the most stable, providing very similar results. Thus we retained the original (and simplest) SLM along with \Eqre{eq:wall-omega}.

The most widely employed closure to model the small scale mixing of the passive scalar in the Lagrangian framework is the interaction by exchange with the mean (IEM) model of \citet{Villermaux_Devillon_72} and \citet{Dopazo_OBrien_74}. This simple and efficient model, however, fails to comply with several physical constraints and desirable properties of an ideal mixing model \citep{Fox_03}.  The interaction by exchange with the conditional mean (IECM) model overcomes some of the difficulties inherent in the IEM model. In this study we justify the use of the IECM model by its being more physical and more accurate, but we acknowledge that it markedly increases the computational cost \citep{Bakosi_08}.

\section{Acknowledgment}
It is a pleasure to ackowledge the fruitful discussions during the course of the present work with Dr.\ Guido Cervone at George Mason University.
\label{sec:acknowledgment}

\bibliographystyle{elsart-harv.bst}
\bibliography{jbakosi}

\end{document}